\documentclass[10pt]{article}

\usepackage[top=0.85in,left=1in,
right=1in,footskip=0.75in]{geometry}
\raggedright
\usepackage{helvet}

\usepackage{latexsym}
\usepackage{epsfig,color}
\usepackage{multirow}
\usepackage{wrapfig}
\usepackage{float}
\usepackage{epstopdf}
\usepackage{mathrsfs}
\usepackage{chemarr}

\usepackage{graphicx}
\usepackage{amsmath}
\usepackage{amssymb}
\usepackage{relsize}
\usepackage{longtable}
\usepackage{multirow}
\usepackage{rotating}

\usepackage{graphicx}
\usepackage{subfig}
\usepackage{multicol}
\usepackage{epstopdf}

\usepackage[pagewise]{lineno}
\usepackage{microtype}
\DisableLigatures[f]{encoding = *, family = * }

\usepackage{rotating}

\usepackage[aboveskip=1pt,labelfont=bf,labelsep=period,justification=raggedright,singlelinecheck=off]{caption}

\makeatletter \renewcommand{\@biblabel}[1]{\quad#1.}  \makeatother

\usepackage{lastpage,fancyhdr,graphicx}
\usepackage{epstopdf}
\def\cO{{\mathcal O}}

\def\G2{{${\rm G}_{\alpha_2\beta\gamma}$}}
\def\cf{{\em cf.\,}}

\newcommand{\x}{{\bf x}}
\newcommand{\y}{{\bf y}}
\newcommand{\mI}{\ensuremath{\bf I}}
 
\newcommand{\bsm}[1]{\boldsymbol{#1}}

\begin{document}
\begin{flushright}{\today}\end{flushright}
\vspace*{0.35in}
\begin{flushleft}
{\Large \textbf\newline {The Role of Cytonemes and Diffusive
  Transport in the Establishment of  Morphogen Gradients } }
\newline
\\Jay Stotsky \textsuperscript{1,*},  Hans Othmer\textsuperscript{1}
\\
\bigskip
{\bf 1} School of Mathematics, University of Minnesota, Minneapolis, MN, USA \\
\bigskip
\end{flushleft}
\tableofcontents
\section*{Abstract}
  
 Spatial distributions of morphogens provide positional information in
 developing systems, but how the distributions are established and maintained
 remains an open problem. Transport by diffusion has been the
 traditional mechanism, but recent experimental work has shown that cells can
 also communicate by filopodia-like structures called cytonemes that make direct
 cell-to-cell contacts.  Here we investigate the roles each may play
 individually in a complex tissue and how they can jointly establish a reliable
 spatial distribution of a morphogen.
  
\section{Introduction and background}

Pattern formation in embryonic development is a major, yet poorly understood,
process in developmental biology.  Throughout early development, maternal and
zygotic cues regulate gene expression, cell proliferation, and differentiation
in space and time in a highly-reproducible manner.  Various theories have been
proposed to explain how individual cells in an aggregate of essentially
identical cells differentiate into a collection of different cell types
organized into the appropriate spatial pattern in response to factors called
morphogens.  The analysis of models of how the spatiotemporal distribution of
morphogens can be controlled so as to produce the appropriate cell type at the
correct location in a tissue is an area in which mathematical modeling has led
to significant insights  \cite{Othmer:2009:ITA}. There are currently two major
theories of morphogen-based pattern formation: Turing's theory
 \cite{Turing:1952:CBM} and the theory of positional information.  In Turing's
theory patterns arise from the interaction of reaction and transport by
diffusion without external cues. This can explain certain patterns such as spots
on insects, but is less useful when the spatial pattern is determined by a
specified distribution of morphogens, and here the theory of positional
information due to Child \cite{Child:1941:PPD} and Wolpert
 \cite{Wolpert:1969:PIS} is more appropriate.

In positional-information mechanisms of morphogen-mediated patterning, cells
determine their spatial position by sensing the local levels of a
spatially-graded morphogen, often produced at the boundary of a region to be
patterned. The simplest model of how this functions is embodied in the French
flag problem, which is to subdivide a finite interval $[0,L]$ on the $x$-axis into
three equal partitions, the first to be colored blue, the second white, and the
third red. This is accomplished by producing a morphogen at the left boundary,
the distribution of which then evolves via diffusive transport and first-order
decay in the interior, and removal at the right boundary. This leads to a
monotone-decreasing gradient of the morphogen level. The interpretation of the
gradient involves two thresholds that then lead to the appropriate coloration of
the interval. This mechanism has gained widespread appeal for interpreting the
appearence of different cell types in an initially-unstructured tissue such as
the {\em Drosophila} wing disc  \cite{Stapornwongkul:2021:GEM}, but it also
raises numerous questions, such as how does the mechanism cope with tissues of
different sizes  \cite{Umulis:2013:MSP}.

\begin{figure}[h!]
\vspace*{-10pt}
  \centerline{
    \includegraphics[width=5.in]{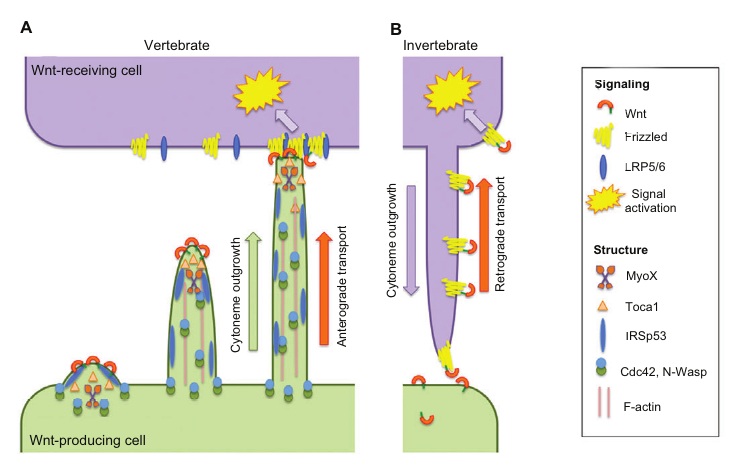} }
\vspace*{20pt} \centerline{
    \includegraphics[width=2.25in]{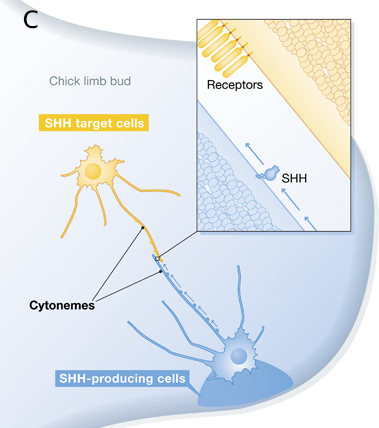} }
\caption{\footnotesize The three modes of cytoneme transport. (A) the PIT mode,
  (B) the RIT mode, (C) the synapse mode. Note that individual cells can give
  rise to multiple cytonemes. From
  \protect \cite{Stanganello:2016:RCW} [(A) and (B)], and
  \protect \cite{Kornberg:2013:CET} (C). }
 \label{PIT-RIT}
\end{figure}

Both Turing's theory and the theory of positional information rely on diffusion
as the means of transport  \cite{Muller:2013:MT,Stapornwongkul:2021:GEM} but
other means of communication in tissues have been discovered in recent years,
and the one of interest here uses what are called cytonemes.  Cytonemes are
rod-like structures similar to filopods, which extend from a cell by
polymerizing actin at the tip. They are $\cO(0.1\!-\!0.5) \mu m$ in diameter and
up to 100 $\mu m$ in length.  In the {\em Drosophila} wing disc they can be as
long as 80 $\mu m$ which spans the entire length of the disc at early stages
 \cite{Ramirez:1999:CCP}. As with filopodia, cytonemes contain an actin network,
but in addition, some contain myosin motors such as Myo-10 that can step along
actin filaments to transport cargo  \cite{Hall:2021:CDS}. Connections between a
cytoneme and a cell, or directly between cytonemes, leads to three forms of
communication between cells, as shown in Figure \ref{PIT-RIT}. The first method,
shown in panel (A), is used in vertebrates to transfer the morphogen Wnt from
producers to receivers, and we call this producer-induced transport (PIT). In
this mode a morphogen-producing cell extends a cytoneme that attaches to a
receiver cell and transfers a bolus of morphogen to the receiver. In some cases
there may be successive transfers of morphogen-filled vesicle-like structures
carried along the cytoneme by molecular motors. In a second mode, (B), the
cytoneme is generated at a non-producing receiver cell such as in the wing disc
of {\em Drosophila}  \cite{Stanganello:2016:RCW} and connects to a producer cell,
from which it either receives a bolus of the morphogen Wg and then retracts, or
receives multiple doses that are transported along the outside of the cytoneme
(see (B)). We call this receiver-initiated transport (RIT).  Finally, in a third
type (C), both receiver and sender send out cytonemes that can create
synapse-like connections to transfer morphogen from producer to receiver. This
occurs in the chick limb bud, as shown in (C). In all cases, individual cells
can employ multiple cytonemes.  One sees that PIT is more direct since the
receiver gets the signal upon contact with the cytoneme, whereas in the RIT mode
the receiver must extend and retract the cytoneme to obtain the signal. Why Wnt
transport in vertebrates uses the former whereas Wg transport in invertebrates
is via the latter remains an open question.

One of the first experimental studies of the role of cytonemes showed how air
sac primordia in the {\em Drosophila} wing disc extend cytonemes to both Fgf-
and Dpp-secreting cells  \cite{Ramirez:1999:CCP}. Recent work shows that they
play a role in a wide variety of systems, including development of neuron types
in the mouse neural tube  \cite{Hall:2021:CDS} and in Hh transport in the wing
disc  \cite{Bischoff:2013:CRE}.  At present the available experimental
information is relatively high-level, demonstrating the existence of cytonemes
and what their cargo is, but details on what controls the origination and number
of cytonemes per cell in  either the PIT or RIT mode remain sparse. Do
extracellular signals bias the search process, or is the process random? How do
cytonemes remain in contact during either loading or unloading of their cargo,
and when vesicles are involved in PIT cytonemes, what determines when the
transfer process stops? A number of recent reviews cover the variety of systems
in which cytonemes play a role
 \cite{Daly:2022:RMC,Korenkova:2020:FIC,Routledge:2019:MIW,Stapornwongkul:2021:GEM,Yamashita:2018:SIC},
but much remains to characterize the details of the processes involved.
   
Nonetheless, a number of models of communication and pattern formation based on
cytonemes have been formulated. One of the first is that due to Vasilopoulos and
Painter  \cite{Vasilopoulos:2016:PFD}, in which the authors develop and analyze a
model for fixed direct contacts between signaling and receiving cells. Their
focus was on the effect of long-range signaling in lateral-inhibition mechanisms
of the type used in Notch-Delta signaling, and in their model the signaling
network is static and a weight function is used to define who gives what and who
gets what. They show that a variety of new pattern types can be generated,
including sparse patterns of isolated cells and large clusters or stripes. Other
models have been based on static or dynamic extension or retraction of cytonemes
and the resulting search process. Similar models in which cytonemes are
permanently attached to cells were analyzed in  \cite{Teimouri:2016:NMU}, wherein
$N$ cytonemes were  statically-attached to cells in a one-dimensional array,
and a deterministic rate of transport was determined by the distance from the
source cell. The authors showed how this could lead to a spatial gradient in the
array. Stochastic transport of packets of morphogen were incorporated in
 \cite{Bressloff:2018:BTM,Kim:2018:DVS}, in which the authors described transport
by a velocity jump process widely used to describe the movement of cells and
organisms  \cite{Othmer:1988:MDB,Hillen:2009:UGP}. Reversal of the direction of
packets was incorporated in the model, but since packets are carried along actin
fibers by myosin motors  \cite{Hall:2021:CDS}, pauses are admissable, but
reversals of direction are not. More recently this model was applied at the
level of a single cytoneme, which can pause and reverse, in a two-dimensional
array of target cells \cite{Bressloff:2019:SCM}. A similar model was used to compare the steady-state
gradient of morphogen resulting from cytoneme transport with that resulting from
diffusion  \cite{Fancher:2020:DVD}, and a  recent computational model was used to
analyze morphogen gradient establishment using cytonemes
 \cite{Aguirre-Tamaral:2021:IUC}.

As we show herein, cytoneme transport is simple and direct, even in complex
tissues, whereas diffusive transport frequently involves intermediate steps such
as binding or cell-to-cell transport (transcytosis). Our first objective is to
develop a population-based PIT model for the cytonemes from which we can predict
the length distribution of cytonemes as a function of space and time, given that
each is initiated at a random time at the source, driven by a Poisson process,
and makes contact for delivery at random lengths (and hence 'age').  Another is
to compare the distribution of morphogen resulting from cytoneme transport with
that which results from using a macroscopic diffusion equation whose
coefficients are based on microscopic properties.

 While we do not cover many possibilities in this article, our goal is
 to lay the groundwork for more detailed cytoneme modeling. Further, more detailed modeling is also difficult because there is
 currently a dearth of experimental evidence,  to understand how hypothetical
 interactions occur. We  believe that the stochastic approach will be
 promising in the future because it is very flexible with regards to what can be
 added or removed from the model. In particular, the stochastic model we will
 introduce is discrete at the level of individual cytonemes and cells, and thus
 can easily incorporate additional effects and interactions as experimental
 evidence accrues towards understanding the behavior of cytonemes.  

\section{Cytoneme-based transport}

We first consider several simple models to show how cytonemes spreading from a
single cell can lead to a graded morphogen distribution. We will assume a
constant cytoneme velocity, and later allow for pausing and other
biologically-relevant processes.

\subsection{Producer-initiated transport}\label{subsec:PIT}
We assume that there is a producer cell located at $x=0$ that intermittently
produces cytonemes at rate $\lambda$ in a one-dimensional domain. The cytonemes
extend at velocity $v$ until they contact a receiver cell at some position $x$,
and release their cargo instantaneously.  The evolution of this system is
governed by
\begin{align}
\begin{aligned}
\dfrac{\partial }{\partial t}p(x,t) + \dfrac{\partial }{\partial x} (v(x)p(x,t))
&= -\mu p(x,t) \\ \dfrac{\partial }{\partial t}M(x,t) &= \gamma\mu p(x,t)
\end{aligned}
\end{align}
where $p(x,t)$ is the number density of cytoneme tips per unit length and $\mu$
is the cytoneme attachment rate at $x$.  In the second equation,  $M$ is the amount of morphogen
delivered per unit length, measured for example in molecules per unit of cell
length, and $\gamma$ is a conversion factor for the amount of morphogen
delivered per cytoneme tip.  We are implicitly assuming that each cytoneme
carries the same amount of cargo and that every attachment is successsful in
delivering morphogen.  We impose a boundary condition at $x=0$ of
the form $p(0,t) = \lambda(t)$, where $\lambda $ is the number of new cytonemes
produced as a function of time. If this equation refers to the position of a
single cytoneme starting at $t=0$, then $\lambda(t)=\delta(t)$. On other hand,
if there is a probability distribution for the emergence of cytonemes, then
$\lambda(t)$ is equal to the rate of that distribution.

The  solution of the first equation can  be found via the method of
characteristics by setting 
\begin{align}
\begin{aligned}
\dfrac{dt}{ds} &= 1 & \quad t(\xi,0) &= \xi \\ 
\dfrac{dx}{ds}&= v & \quad x(\xi,0) &= 0\\ 
\dfrac{dp}{ds} &= -\mu p& \quad p(\xi,0) &= \lambda(\xi),
\end{aligned}
\end{align}
where $s\ge 0$ parameterizes a characteristic and $\xi$ parameterizes the curve 
$x = 0$. One finds that the  solution is
\[
p(x,t) = \lambda(t) H\left(t-\frac{x}{v} \right)e^{-\frac{\mu}{v}x}.
\]
where $H$ is the Heaviside function.  While this model is very simple, it is
already clear that an important factor in the morphogen distribution is the
ratio $\mu/v$, which essentially governs how quickly cytonemes halt compared
with their extensional velocity. If this quantity is large, the steady-state
morphogen distribution is highly concentrated at $x=0$, whereas it is
widely-distributed when it is small. The dynamic approach to the exponential
distribution in space behind the front  $x = vt$ is governed solely by the
velocity - more rapid spreading occurs with a larger velocity - as is expected.

The distribution of morphogen received by the cells  is then
given by
\[
M(x,t) = \gamma\int_0^t \mu p(x,\tau)d\tau =\gamma\mu e^{-\frac{\mu}{v}x}\int_{x/v}^t
\lambda(z)dz
\]
For $\lambda(t) = \lambda_0H(t)$ where $H(t)$ is the heaviside function, the result is 
\[
M(x,t) = \gamma\mu\lambda_0H\left(t-\frac{x}{v}\right)e^{-\frac{\mu}{v}x}\left(t-\frac{x}{v}\right)
\]
Noteworthy is the fact that there is a time-delay after
startup during which $M(x,t)$ is zero. This is in contrast to diffusive
mechanisms, which instantaneously generate nonzero concentrations
globally. Also note that for each $x$, once $t>x/v$, the concentration increases
linearly in time at that point, with a coefficient dependendent on $x$. These results
are shown in Figure \ref{fig:PIT-Morphogen}.\footnote{In this and the
  following figures we use dimensionless variables based on a time scale of 10
  seconds and a length scale of 10 microns.}

\begin{figure}[h!]
\begin{center}
\includegraphics[width = 0.45\textwidth]{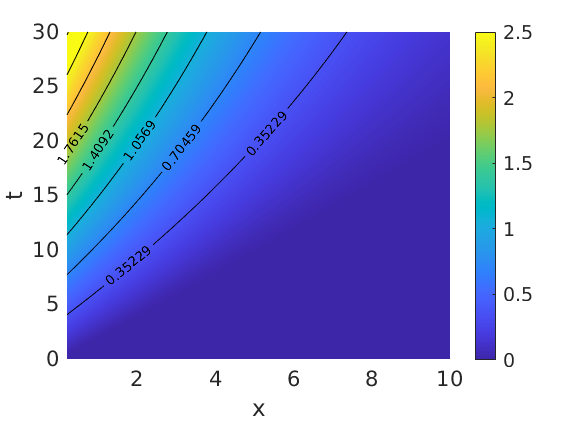}
\includegraphics[width = 0.45\textwidth]{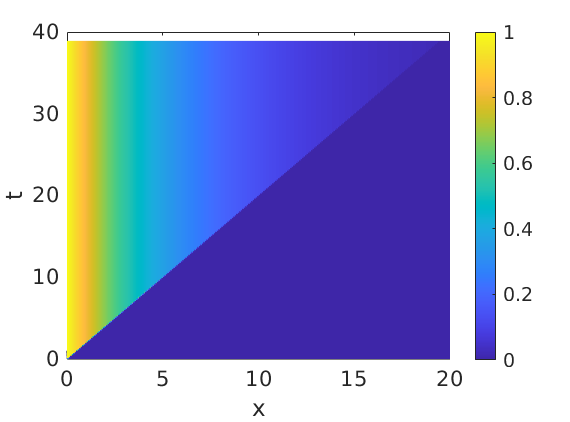}
\caption{ The morphogen concentration  $M(x,t)$ as a function of $x$ and $t$ (left)
   and  $p(x,t)$ (right) for the  PIT model. 
 The parameter values used are $\mu_d = 0.1, v = 0.5$ and $\lambda = 1$.}
\label{fig:PIT-Morphogen}
\end{center}
\end{figure}
While there are many more details that can be analyzed  for the PIT
process, we next consider the RIT process in greater detail.

\subsection{Receiver-initiated transport}
  Under RIT non-producers search for producers to extract a bolus of
  morphogen. This mode is interesting because the cargo has direct access to the
  cytosol, whereas in PIT the transfer is thought to involve surface
  receptors. Thus there may be local loss of signals if the bolus saturates the
  available receptors. When the cargo in RIT is a morphogen, it may bind to
  receptors in the cytosol to activate downstream signaling.

Suppose that when a cytoneme makes contact with a producer cell it loads its
cargo, which could be a morphogen in a vesicle, and retracts.  The cytoneme
extension process can be modeled as above, except that the position of
the receiver cell is now very important, and the stopping rate $\mu$ must now
be spatially variable since cytonemes that stop and perhaps retract  prior to reaching any producing
cells would be ineffective.  In particular, we fix the receiver cell at position
$x_0>0$ and assume that the producer cells are located at $x=0$.  We further
assume that attachment occurs instantaneously once a cytoneme reaches $x=0$.
Then the number density per unit length of extending cytoneme tips is governed by
\begin{align}
\begin{aligned}
\frac{\partial p_{e}}{\partial t}-v\frac{\partial p_{e}}{\partial x} &= 0,
\ \ \ \ x\in(0,x_0] \\ p_e(x_0,t) &= \lambda(x_0,t)
\label{eq:Extend}
\end{aligned}
\end{align}
where $\lambda(x_0,t)$ is the number density of cytonemes generated by the
receiver cell.\footnote{In proposing this form of equation, we implicitly assume
several things. Firstly, we assume that only the cytoneme tip contains
morphogen. Thus we only need to track the tip position rather than the entire
shape of the cytoneme.  In addition, we assume that space-exclusion effects are
negligible, which allows us to assume that each cytoneme moves independently of
the others.  The rationale behind these assumptions is that the cytonemes are
quite small, and thus, even if there were many cytonemes in a small area, there
would still be sufficient space for the cytonemes to wriggle through without
strong interaction effects. On the other hand, it is important to note that
there is a dearth of biological evidence for how cytonemes actually move through
a tissue and whether they interact with one another or not. An example of
interaction is shown in Figure \ref{PIT-RIT}(c), but this is between two
different types of cytonemes. The effect of interactions between like types
remains for future work, and will also hinge upon future experimental
observations. It should also be noted that we do not track cytonemes that do not
reach $x = 0$ since they do not contribute to the morphogen flux. To account for
them one could include a 'death' term, but we ignore this.}

The number of cytonemes  per unit of time that reach producer cells at $x=0$ is
$J(t)=vp_e(0,t)$, and the number  of attached cytonemes, $P_a$ evolves 
according to 
\begin{equation}
\frac{dP_a}{dt} = J(t)-\mu_dP_a, 
\label{eq:Attached}
\end{equation} 
where $\mu_d$ is the detachment rate. 
After  detachment, the cytoneme is presumed to be carrying a morphogen
packet, and the density of cytoneme tip positions $p_r$ for  cytonemes
laden with morphogen is governed by\footnote{If there is a probability of an attached cytoneme failing to obtain a morphogen packet, we can scale the boundary condition on $p_e$ by the probability of success to obtain the population of morphogen-laden cytonemes.}
\begin{align}
\begin{aligned}
\frac{\partial p_{r}}{\partial t}+v\frac{\partial p_{r}}{\partial x} &= 0,
\ \ \ \ x\in [0,x_0) \\ vp_r(0,t) &= \mu_dP_a.
\end{aligned}
\label{eq:Retract}
\end{align}
This system can be solved analytically once $\lambda(x_0,t)$ is specified. In
particular, let us assume that a cell turns on a constant rate $\lambda_0$ of
cytoneme generation at $t=0$ , thus $\lambda(x_0,t) = \lambda_0 H(t)$ where
$H(t)$ is the Heaviside step function.  The solution is then

\begin{align}
\begin{aligned}
p_{e}(x,t) &= \lambda_0H\left(t-\frac{x_0-x}{v}\right)\\ 
P_a(t) &= \frac{\lambda_0v}{\mu_d} H\left(t-\frac{x_0}{v} \right)\left(1-e^{-\mu_d\left(t-\frac{x_0}{v} \right)}
\right)\\  
p_r(x,t) &= \lambda_0 H\left( t-\frac{x_0+x}{v}\right)\left(1-e^{-\mu_d\left(t-\frac{x_0+x}{v}  \right)}\right) \\ 
M(t,x_0) &= \gamma v\lambda_0 H\left(t-\frac{2x_0}{v}  \right)\left[t-\frac{2x_0}{v} +\frac{1}{\mu_d}\left(1-e^{-\mu_d\left(t-\frac{2x_0}{v}  \right) } \right) \right]
\end{aligned}
\label{advection}
\end{align}

\noindent
As before,   the constant  $\gamma$ gives the amount of morphogen transferred per cytoneme, and
$M$ is the morphogen received by a cell at $x_0$.

Due to the finite velocity, we see that at  there is an initial  time interval 
$(0,2x_0/v)$ during which  no morphogen is received, followed by gradual morphogen
reception that depends on the attachment waiting-time distribution,
and this gradually ramps up to a linear rate of increase with slope
$v\gamma\lambda_0$.  Note that this linear increase rate is independent of $x_0$,
but that the time-delay required to reach that rate is $x_0$-dependent.

If there are many cells, each at different points, $x_0$, then the overall
cytoneme tip distribution can be found by considering the distributions ($p_e$,
$P_a$ and $p_r$ above) as functions of $x_0$ and summing over $x_0$. The
concentration of morphogen in receiver cells will then increase in a wave-like
fashion across the tissue due to the variable time-delays. However, if all cells
have the same $\lambda_0$ and $v$, the concentration will approach a linearly decreasing function of $x_0$
after the exponential terms relax. This is in contrast to diffusion with decay, where the
profile develops into an  exponential.

However, this  model is too simplistic, because cytonemes do not
simply extend for indefinite lengths of time at a particular velocity, but are
known to exhibit stopping and starting. In fact, the time spent resting is known
to exhibit Poissonian statistics. Furthermore, some cytonemes may fail to reach
their target and retract prior to receiving a morphogen packet. To account for
these effects, we include decay terms and first-order reaction terms in the
cytoneme extension and retraction equations, and introduce ordinary
differential equations to describe the population of cytonemes in a resting
phase. This leads to the following system, 
\begin{align}
\begin{aligned}
\frac{\partial p_{e}}{\partial t}-v\frac{\partial p_{e}}{\partial x} &=
-\lambda_r p_e + \lambda_m r_1 - k_d p_e, & p_e(x_0,t) = \lambda(t), &\quad \mbox{extension}
\\ 
\frac{dr_1}{dt} &= \lambda_r p_e - \lambda_m r_1, &  & \quad \mbox{rest phase} \\
 \frac{dP_a}{dt} &= vp_e(0,t) - \mu_d P_a, & &\quad \mbox{attachment} \\
\frac{\partial p_{r}}{\partial t}+v\frac{\partial
  p_{r}}{\partial x} &= -\lambda_r p_r + \lambda_m r_2, & vp_r(0,t) = \mu_d
P_a, & \quad \mbox{retraction} \\
\frac{dr_2}{dt} &= \lambda_r p_r -\lambda_m r_2, & &\quad \mbox{rest phase}
\\ \frac{dM}{dt} &= \gamma
vp_r(x_0,t), & & \quad \mbox{morphogen accumulation}
\end{aligned} 
\label{eq:RestingPhaseModel}
\end{align}
where again, the domain is $(0,x_0)$. Here $r_1$ and $r_2$ are two distinct
resting phases for the extending and retracting cytonemes.  It is a minor
technical complication that these two populations must be tracked separately,
because otherwise we would lose track of which cytonemes in the resting phase
contain morphogen and which do not. Note that the cytonemes that reverse prior
to reaching the source region are not included in $p_r$ since they do not
contribute to the morphogen flux, but rather are simply eliminated via the
degradation term $k_d p_e$.

\begin{figure}[h!]
\begin{center}
\includegraphics[width = 0.475\textwidth]{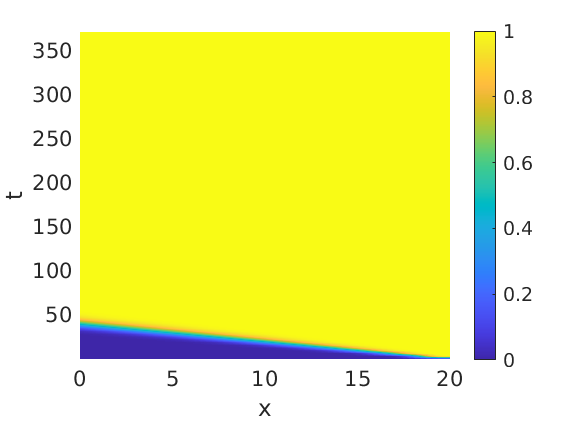}
\includegraphics[width = 0.475\textwidth]{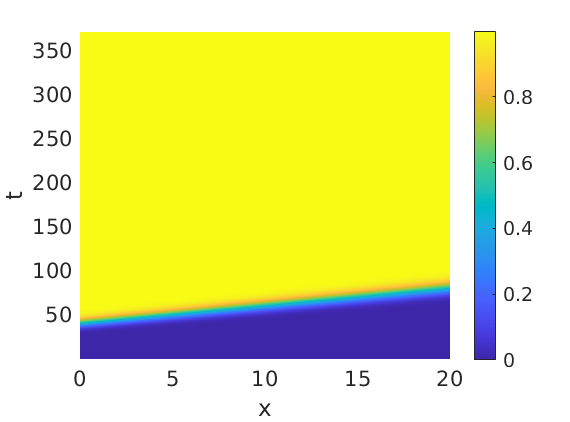} \\
\includegraphics[width = 0.475\textwidth]{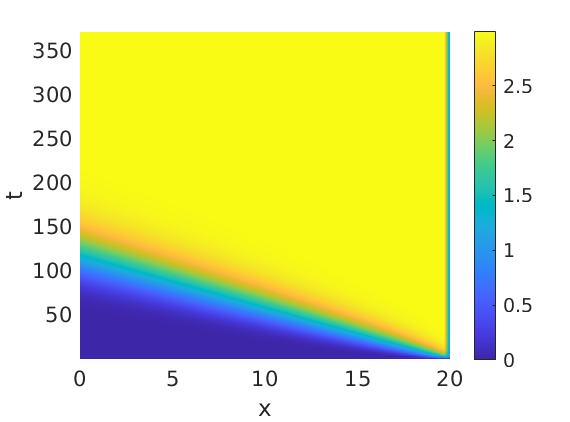}
\includegraphics[width = 0.475\textwidth]{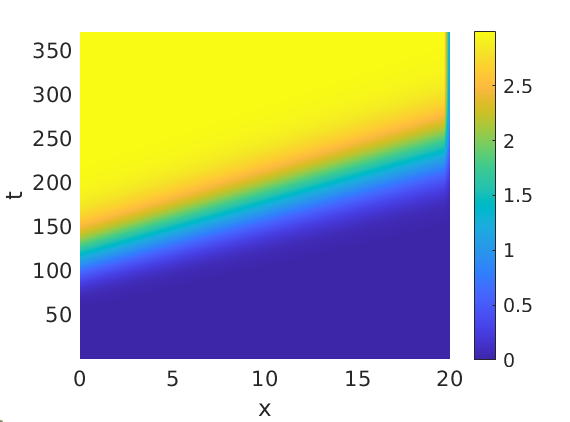} 
\end{center}
\caption{Plots of the space-time cytoneme tip location distribution -- computed
  from Equation \ref{advection} -- for cytonemes emanating from a cell at
  $x_0=20$. The top row is the pure advection case ($\lambda_r
  =0$), with $p_e$ on the left and $p_r$ on the right. The bottom row includes a
  rest phase, with  $p_e(x,t)+r_1(x,t)$ on the left and $p_r(x,t)+r_2(x,t)$ on
  the right. We set $\mu_d = 1s^{-1}$, $\lambda_m=0.25s^{-1}$, $\lambda_r =
  0.5s^{-1}$, and $v=1\mu m/s$.
\label{fig:P_vs_x_t} }
\end{figure}

The resulting system can be
solved via Laplace transform, but the inversion integral appears to be
non-elementary. Thus, we have solved the above system numerically using an upwind finite difference approximation to obtain
results for $M(t)$ and the other quantities of interest. In comparison with the
simpler model, the main additional quantities to study here are the decay rate
$k_d$ of cytonemes that fail to obtain morphogen, and the rates of state changes
between resting and mobile states. The spatial distributions of the cytonemes for a
cell fixed at $x_0=20$ are shown in Figure \ref{fig:P_vs_x_t}. 

In that figure the value of $p(x,t) = p_e(x,t)+r_1(x,t)$ (the density of
cytonemes in $(x,x+dx)$ at $t$) becomes larger when the average length of the
resting phase increases. This is because as the average resting time increases,
the cytoneme transport process is less efficient, and thus the time-average
velocity of cytonemes decreases. At the same time, the cytoneme generation rate
is not affected by the resting phase in our model, and thus on average, the same
number of cytonemes are still generated in each time interval. Since cytonemes
are not traveling as quickly, but are still generated at the same rate, the
spatial density of cytonemes must then be higher in the absence of degradation.
Also note that the profiles of extending and retreating cytonemes are
qualitatively-inverted, as expected.  With degradation, the degradation rate
also figures into the balance, with a higher degradation rate leading to lower
cytoneme densities.

Since the resting phase decreases the efficiency of cytoneme transport, one sees
that although the number of cytonemes that exist simultaneously may be much
larger than when there is no rest phase, the amount of morphogen delivered is
largest  when there is no resting phase. This leads to the question of why cytonemes exhibit a resting phase at all - it seems to only hinder the morphogen delivery
process. While it is difficult to say precisely why cytonemes have resting
phases, it may be that without biological regulation, the resting
phases may be much longer than those observed, and that in fact, cytoneme
transport processes have been fine-tuned to minimize the resting phase. Another
aspect may be that working in one spatial dimension, the problem of orienting the cytoneme is not as significant as it would be for a cytoneme extending in
a 3D or 2D tissue. It may be that during the resting phases, the cytoneme is
integrating various guiding queues to ensure that it travels towards its
target. Extending the model to include various aspects of cytoneme
transport is an area of future interest, though we do make some initial inroads
into the importance of cytoneme transport direction in the stochastic model
described in the next section.

Before delving into a stochastic model, we conclude this section by describing 
several extensions that can be easily-included  in the continuum model.  In
the models described above, we have conditioned on the receiver cell being
located at some point, $x_0$. However, it is straightforward to allow various
starting points and then compute $M(t,x_0)$ to obtain the spatiotemporal
dependence of the morphogen concentration in the receiver cells.

Secondly, when  there is
cytoneme degradation, the resulting concentration profile exhibits 
exponential decay  rather than linear decrease.
To understand this, one can think of the cytoneme extension process as a
discrete random walk problem in which the cytoneme can either move forward or
permanently stop at any time. Let $\alpha$ be the halting probability for a
single step. Then, after $n$ steps, the probability that a particular cytoneme
has reached $x_n = n\delta x$ is $(1-\alpha)^n$.  Thus, if we consider a new
cytoneme being added at each step at $x_0=0$, we find  that after sufficient  time,
the probability of a cytoneme  at $x_k$ is $p(x_k) = (1-\alpha)^k$. With $x_k =
k/n$, and $\alpha_n = \alpha/n$, we hold $k/n$ fixed and allow  $n\rightarrow
\infty$. In this limit, we obtain an exponential decay for the density of
cytonemes in $(x,x+dx)$, because  $\lim_{n\rightarrow\infty} (1-\alpha_n)^{xn} =
e^{-\alpha x}$.

\begin{figure}
\centerline{
\includegraphics[width = 0.45\textwidth]{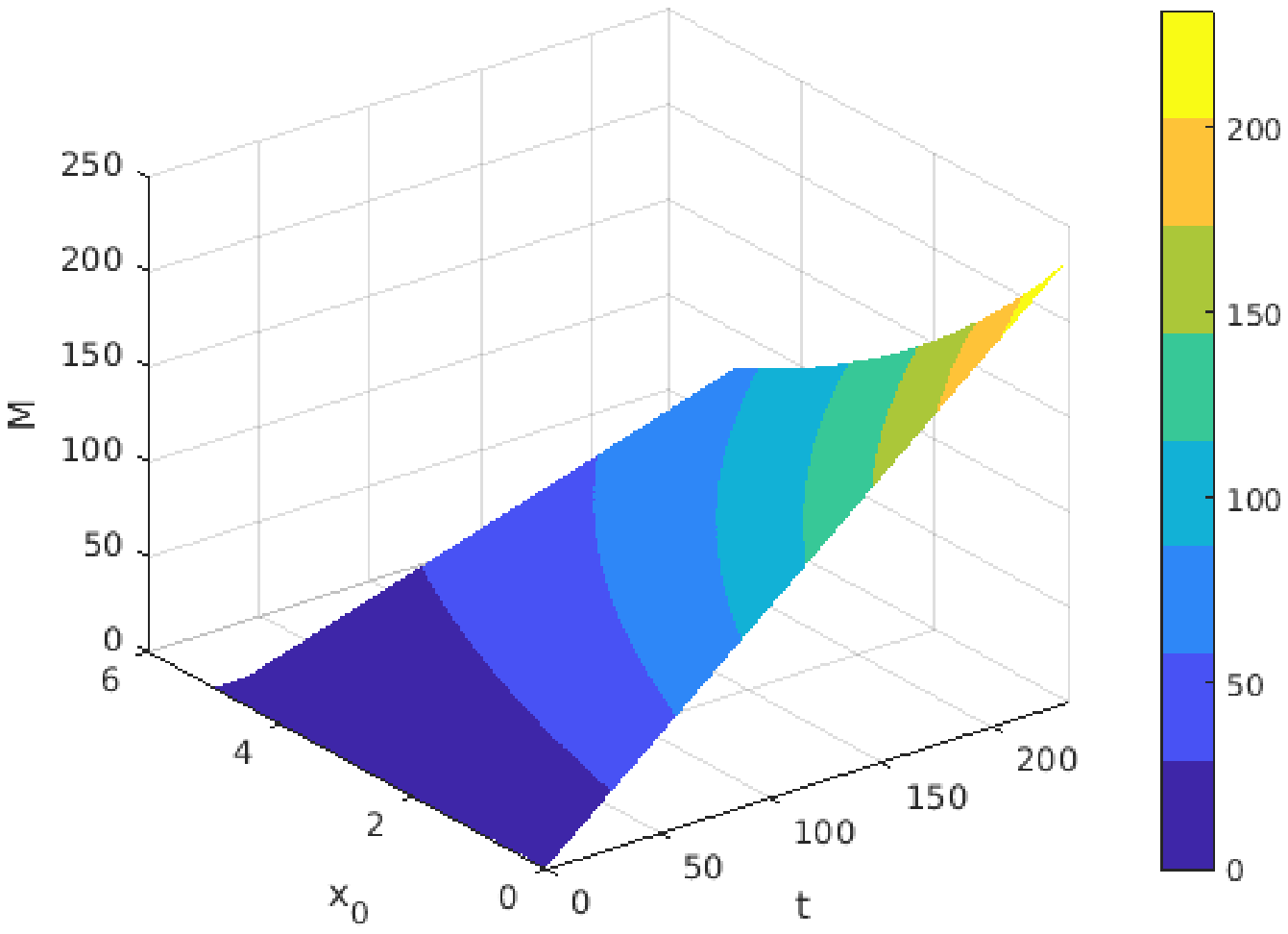}
\includegraphics[width = 0.45\textwidth]{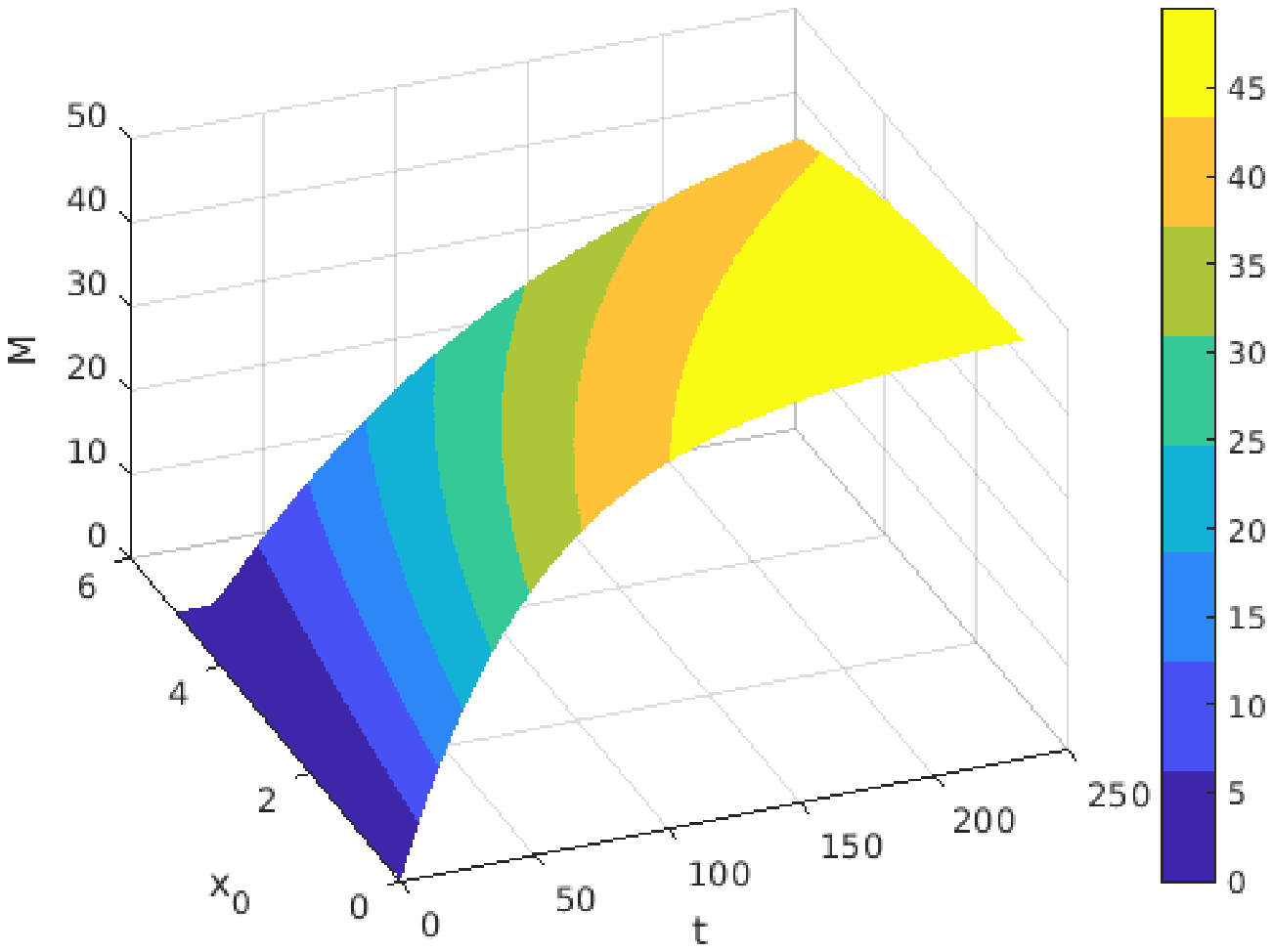}
}
\caption{ The evolving  number of morphogen
  packets $M(t)$ received as a function of time and
  space. On the left there is no cellular feedback, so the cells continue 
  receiving morphogen, whereas on the right, there is feedback that shuts of cytoneme
  generation after 50 packets are received. The bright yellow shows  where $M$
  reaches its maximum, and one can see that this peak gradually expands outwards
  towards more distant cells as time progresses.  
  }
\label{fig:M_Wave}
\end{figure}

Further details can be included if we allow receiver cells to alter the cytoneme
production rate $p_e(x_0,t)$ in response to the amount of morphogen received. In
this case  the model becomes nonlinear, and cannot be solved
analytically. Morphogen feedback might be expected biologically if we presume
that the primary role of the morphogen is to activate some signalling cascade
involved in cell differentiation. Once the appropriate level of the  signal has been received,
the cell would no longer require additional morphogen, and thus would not need
to continue generating cytonemes. In this case, we set $\lambda(t) =
\lambda(M(t))$ to be a decreasing function of $M$. In the simplest case,
$\lambda = \lambda_0 - \lambda_1 q$ is a linearly decreasing function. The ratio
$\lambda_0/\lambda_1$ determines the maximum amount of morphogen the cell will
receive before it shuts off cytoneme generation. This leads to a wave of cells
receiving a particular amount of morphogen, approximately $M =
\lambda_0/\lambda_1$, as is shown in Figure \ref{fig:M_Wave}.

Notice
  that since it takes distant receiver cells longer to reach the producers
  and return with morphogen, there is a lag between the morphogen concentration
  at more distant cells, and the morphogen concentration at the cells with $x_0$
  near $0$ (which is where the producer cells are located).
Interestingly, if cells are able to shut off their cytoneme generation after
sufficient morphogen has been received, and also only grow in response to
adequate morphogen, this may yield a mechanism by which tissue size can  be
controlled.

\subsection{Cytoneme return rates and first-passage-time distributions}

While the preceeding results are based on a continuum description, assuming
sufficiently- large population of cytonemes existing around each point in space
and time to validate the continuum hypothesis, there is also a close connection
between the previous models, and certain stochastic cytoneme models. In
particular, let us assume that a single cytoneme undergoes alternating phases of
extension (or retraction) and resting. The time intervals for each of these
mobile or rest phases are presumed to be random variables that are
Poisson-distributed with rates $\lambda_m$ and $\lambda_r$. In other words, for
a mobile phase, the probability that that particular mobile phase exists for a
time $T>t$ is
\[
\mathbb{P}\left[T>t\right] = e^{-\lambda_m t}
\]
and likewise for the rest phases with $\lambda_m$ replaced by
$\lambda_r$. Similarly, we have assumed that the attachment phase lasts for a
Poisson distributed time with rate $\mu_d$. One can show via the methods in
\cite{Stotsky:2021:RWA} that with all of the distributions exponentially
distributed, the resulting probabilities $p_e(x,t)$, $P_a(t)$, and $p_r(x,t)$
obey the same PDEs as in Equation \eqref{eq:RestingPhaseModel}, but with
$k_d=0$, and initial condition of the form $p_e(x,0) = \delta(x-x_0)$ that
describes the initial (deterministic) placement of the cytoneme at $t=0$. This
is essentially a consequence of the fact that for i.i.d. random variables, the
expectation of the mean is equal to the expectation of a single instance.

When considering just a single cytoneme moving according to a stochastic
process, a particularly important quantity is the distribution characterizing
the time it takes for a cytoneme to extend, upload  some morphogen
molecules or morphogen-containing vesicles, and return back the the receiver
cell to deliver the morphogen. This probability distribution can be written as  
\[
\mathbb{P}\left[\mbox{return by } t \mbox{ after departing at } t_0 \mbox{ from } x_0\right] \equiv F(t|t_0,x_0) = \int_0^t f(t|x_0,t_0)dt
\]
where $f(t|x_0,t_0)dt$ is the probability for a cytoneme, originating at $t_0$
from a cell at $x_0$ returning back to the cell with morphogen in the interval
$(t,t+dt)$. The return event is an example of a first-passage-time event.  Thus,
$F(t|t_0,x_0)$ represents the total probability that a cytoneme has returned by
time $t$,  which is  the cumulative distribution for the first-passage-time
(FPT). FPT's have been applied to cytoneme-mediated transport previously in
\cite{Bressloff:2019:SCM}, and we will consider several  simple examples
here.  

First, recall Equations \eqref{advection} where the extension ($p_e$),
retraction ($p_r$), and attachment ($P_a$) probabilities were obtained for a
cell generating cytonemes at a fixed rate starting at $t=0$. Iif instead we
condition the process on a cytoneme being initiated at $t=0$ in that example, we
obtain
\begin{align}
\begin{aligned}
p_e(x,t) &= \delta\left(t-\frac{x_0-x}{v}\right)\\
P_a(t) &= \frac{v}{\mu_d}H\left(t-\frac{x_0}{v}\right)e^{-\mu_d\left(t-\frac{x_0}{v}\right)}\\
p_r(x,t) &= H\left(t-\frac{x_0+x}{v}\right) e^{-\mu_d\left(t-\frac{x_0}{v}\right)}.
\end{aligned}
\end{align}

To compute the FPT distribution, we consider the fact that  the probability of
a cytoneme passing the point $x=x_0$ in the interval $(t,t+dt)$ is equal to
$vp_r(x_0,t)$, and that this is equivalent in the limit as $dt\rightarrow 0$ to
the cytoneme experiencing a first-passage event at $t$. Thus, we have that
$f(t|x_0,t_0) = vp_r(x_0,t-t_0)$ for any $t>t_0$. The amount of morphogen
received by the cell is then proportional to $F(t|x_0,t_0)$ which is just the
integral of $f(t|x_0,t_0)$ from $0$ to $t$.  

As a slight modification, the FPT distributions may be allowed to be degenerate
(integrate to some number $0<\alpha<1$) since some cytonemes may fail to ever
upload morphogens, and simply return empty-handed. To see an example of how this
may arise, consider a model where cytonemes extend at a velocity $v$ as before,
but have a probability $\beta$ per unit time of halting and turning back
regardless of whether it has reached the producer cells. Thus, the governing
equations are those in Equations \eqref{eq:Extend} through \eqref{eq:Retract},
but now we modify Equation \eqref{eq:Extend} to become 
\[
\frac{\partial p_e}{\partial t}-v\frac{\partial p_e}{\partial x} = -\beta p_e
\]
The solution, which again  may be obtained via the method of characteristics, is
\begin{align}
\begin{aligned}
p_e(x,t) &= \delta\left(t-\frac{x_0-x}{v}\right)e^{-\beta x/v} \\
P_a(t) &= \frac{v}{\mu_d}H\left(t-\frac{x_0}{v}\right)e^{-\mu_d\left(t-\frac{x_0}{v}\right)-\beta x_0/v}\\
p_r(x,t) &= H\left(t-\frac{x_0+x}{v}\right) e^{-\mu_d\left(t-\frac{x_0}{v}\right)-\beta x_0/v}.
\end{aligned}
\end{align}
The FPT density is as before, but now multiplied by a factor $\pi_0 = e^{-\beta
  x_0/v}$. This factor is called a splitting probability and we can thus write
the FPT density as $\pi_0(x_0) f(t|x_0,t_0)$ where $f$ is unitary.   

We also consider a more complex example in which the cytonemes extend, but can
rest for intermittent intervals of time. Thus, they alternate between motile
phases and rest phases. As in \cite{ricca2010stepping}, we assume that both the
motile and resting phases obey Poisson distributions, albeit with different
rates, denoted $\mu_m$ and $\mu_r$.   This stochastic process is an example of an
alternating renewal process, and these have been analyzed in
\cite{ross1996stochastic,Cox1962Renewal,Bressloff:2019:SCM}. While non-trivial,
the exact FPT for a cytoneme obeying this process is computable via
Laplace-transform methods.

 The first-passage-time distribution can be computed via similar techniques to
 those in  \cite{masoliver1993solution,Bressloff:2019:SCM}, and the details are
 given in the appendix. We find that 
\[
f(t|x_0) = \int_0^t \left[\mu_d e^{-\mu_d(t-\tau)}H\left(\tau-\frac{2x_0}{v}\right)\left[e^{-\left(\tau+\frac{x_0}{v}\right)\lambda_m} \sqrt{\frac{c}{\left(\tau-\frac{2x_0}{v}\right)}} I_1\left(2\sqrt{c \left(\tau-\frac{2x_0}{v}\right) }\right)+\delta(\tau-\frac{2x_0}{v})\right]\right] d\tau 
\]
with $c=2\lambda_m\lambda_r x_0/v$ and $I_1(z)$ is the modified Bessel-function
of order 1. This distribution is quite complex, but the moments of the
distribution can be found easily by differentiating the Laplace transformed FPT(which
is comparatively simpler) to obtain
\[
\langle t\rangle = \frac{1}{\mu_d} + \frac{2x_0}{v}\left(1+\frac{\lambda_m}{\lambda_r}\right)
\]
Notice that the two terms, $1/\mu_d$ and the velocity dependent term essentially
account for  1) the time the cytoneme is attached to the producer cell, and 2) the time
it takes to go out and come back. 

The variance can be computed by taking the second derivative of the Laplace-transformed FPT at $s=0$ and subtracting $\langle t\rangle^2$. This leads to 
\[
\sigma^2(t) = \langle \left(t-\langle t\rangle \right)^2\rangle = \frac{1}{\mu_d^2}+\frac{4x_0}{v}\frac{\lambda_m}{\lambda_r^2}
\]
which again, has components due to the variance of waiting while attached, and
of the time it takes to travel to and from the producer cells.  
Note that in contrast to other models \cite{Bressloff:2019:SCM}, the cytonemes
do not have periods of forward and backward motion while extending. Rather, they
have periods of forward motion with interspersed pauses with no motion. We
believe this more realistically captures the dynamics of cytonemes which we
expect exhibit highly biased transport towards a target.  Nonetheless,  the two models are  similar, and setting $v_-=0$ in \cite{Bressloff:2019:SCM} should yield results very similar to ours here.

\section{Stochastic models of cytoneme-based  transport}
In the continuum models introduced thus far, each quantity refers to an average
number of morphogen packets, but because individual cells only maintain a small
number of cytonemes at any moment, it is important to consider other quantities
such as the variance in the number of morphogen packets received during 
transport. This has further relevance since it appears that in many cases
morphogen delivery is not a continuous process, but coincides with the arrival
of discrete punctae, thought to be vesicles containing high levels of
morphogen. This would be expected to produce bursts of morphogen at discrete
times, separated by intervals where no new morphogen arrives.  To understand
such quantities, further details of the stochastic nature of the cytoneme generation and morphogen
delivery must be taken into account, and we address this in this section.

\subsection[Multiple cytonemes]{Multiple cytonemes emanating from a  receiver cell}

We consider a single receiver cell located at $x_0$ that extends cytonemes to
reach source cells located at $x=0$ as above, but now we analyze the more
realistic situation in which the cytoneme generation rate is stochastic.  In
general, a complete analysis is too complex, but some analytical results can be
obtained if cytoneme generation follows a Poisson process, which implies that
there is no correlation between the generation of one cytoneme, and the time at
which a subsequent one is generated. Rather, there is simply a smoothly varying
rate (or intensity) $\lambda(t)$ such that $\int_{t_1}^{t_2}\lambda(t)dt$ gives
the expected number of new cytonemes generated in $(t_1,t_2)$. This description
is relevant in reality since individual cells can support multiple cytonemes,
and these cytonemes may form in different regions of the cell, thus removing
correlations (or anti-correlations) between the times for subsequent cytonemes
to be generated. The Poisson process approach also seems reasonable for
describing the generation of cytonemes amongst an array of cells as long as
cytoneme generation in a given cell does not significantly alter the cytoneme
generation capabilities of its neighbors.

Thus let $X(t)$ be a Poisson process of rate $\lambda$, and let $E_1, E_2,
\cdots E_n$ be the events of generating cytonemes in $(0,T]$. We seek the joint
distribution of the random times of the ${E_i}$ conditioned on the event
$N(T)=n$.  The probability $P(N(T) = n)$ for a non-homogeneous Poisson process
of intensity $\lambda(t)$ is given by
\[
P(N(T) = n) \equiv p_n (T)= \frac{\left(\int_0^T\lambda(\tau)d\tau\right)^n}{n!}e^{-\int_0^T\lambda(\tau)d\tau},
\]
while for a time-homogneous Poisson process  $\lambda$ is constant, and
\[
p_n(T) = \frac{(\lambda T)^n}{n!}e^{-\lambda T}.
\]

 We can also compute quantities such as the expectation and variance of the
 cytoneme distributions as follows.  For instance, let $x(t|t_i)$ be the
 position at time $t$ of a cytoneme tip that originated at the event time
 $t_i$. If the cytoneme transport is simply advective, then this is equal to
 $x_0 - v(t-t_i)_+$ where $u_+ = max (0,u)$ . Then, since $\{t_i\}$ is IID ,
 given a fixed number $N(t)=n$ of cytonemes the expectation of the average
 position is simply the expected position of any individual cytoneme\footnote{We
   use the notation $\mathbb{E}_n[\cdot]$ to indicate expectations that are
   conditional on $N(t)=n$}
\[
\mathbb{E}_n\left[\frac{1}{n}\sum_{i=1}^nx(t|t_i)\right] = \mathbb{E}_1[x(t|\tau)] = x_0 - \frac{vt}{2}.
\]
Since this result is independent of $n$, it also gives the mean position when
$N$ is allowed to be a random variable. More generally, for some quantity
$Q(t|t_i)$ that depends only on a single $t_i$ from an IID set, the expectation
of the average over any point process is simply the expectation of $Q(t|\tau)$,
\begin{equation}
\mathbb{E}\left[\frac{1}{N(t)}\sum_{i=1}^{N(t)}Q(t|t_i) \right] = \mathbb{E}_1\left[Q(t|\tau)\right].\label{eq:Mean}
\end{equation}
On the other hand, the expectation of the sum $\sum_{i=1}^{N(t)}Q(t|t_i)$ is 
\begin{equation}
\mathbb{E}\left[\sum_{i=1}^{N(t)}Q(t|t_i)  \right] = \mathbb{E}[N(t)]\mathbb{E}_1[Q(t|\tau)] = (\lambda t) \mathbb{E}_1[Q(t|\tau)].
\label{eq:ExpSum}
\end{equation}
Higher order moments such as the variance can be found as follows. We write
\[
\sigma^2\left(\frac{1}{N(t)}\sum_{i=1}^{N(t)} x(t|t_i) \right) = \sum_{n=0}^{\infty} p_n(t)\mathbb{E}_n\left[\left(\frac{1}{n}\sum_{i=1}^nx(t|t_i)-\mathbb{E}_1[x(t|\cdot)]  \right)^2\right]
\]
To evaluate this we split the squared-sum in the formula above into a sum
of diagonal terms  $i=j$ and an 'off-diagonal' term with $i\neq j$. After simplification,
this yields\footnote{To see the last simplification below, recall that
  $\sum_{n=0}^{\infty} \frac{z^n}{n!} = e^{z}$, and that the integral $\int z^n
  dz = \frac{1}{n+1}z^{n+1}$. Thus, formally, we can consider
  $\sum_{n=1}^{\infty} \frac{1}{n}\frac{z^n}{n!} = \int\sum_{n=1}^{\infty}
  \frac{z^{n-1}}{n!}dz = \int \frac{e^z-1}{z}dz$. The resulting exponential
  integral is a transcendental function and cannot be further reduced into a
  finite combination of other elementary functions.  },  
\begin{align}
\begin{aligned}
\sigma^2(x) &= \sum_{n=0}^{\infty}\frac{(\lambda t)^n}{n!}e^{-\lambda t}\left[\frac{1}{n}\mathbb{E}_1\left( x(t|\tau)^2\right) +\frac{n-1}{n}\left(\mathbb{E}_1 x(t|\tau)\right)^2\right] - \left(x_0 - \frac{vt}{2}\right)^2\\
&=\sum_{n=0}^{\infty}\frac{(\lambda t)^n}{n!}e^{-\lambda t}
\frac{1}{n}\sigma^2_1(x(t|\tau) = \frac{(vt)^2}{12}e^{-\lambda t}\int_0^{\lambda
  t}\frac{1-e^{-z}}{z}dz.
\end{aligned}
\label{eq:Var}
\end{align}
Note that the $x_0$ terms all drop out of this expression. This is due to the fact that $x_0$ is fixed in this calculation, and thus the variance is in the cytoneme position is independent of $x_0$. 

More generally,  $Q(t|t_i)$  could be   more complicated, such as the
probability of the cytoneme having returned to the starting point or reached the
source cells.  Under the above conditions the  variance of the average  or the total
sum of $Q(t|\cdot)$, given by 
\[
\sigma^2\left(\frac{1}{N(t)}\sum_{i=1}^{N(t)}Q(t|t_i)\right)
\ \ \ \ \ \ \mbox{or} \ \ \ \ \sigma^2\left(\sum_{i=1}^{N(t)}Q(t|t_i) \right)
\]
can then be computed.  For the mean, $\frac{1}{N(t)}\sum Q(t|t_i)$, the variance
of the mean is found by generalizing Equation \eqref{eq:Var} as follows 
\begin{equation}
\sigma^2\left(\frac{1}{N(t)}\sum_{i=1}^{N(t)}Q(t|t_i) \right) = \sigma^2_1(Q(t|\tau))e^{-\lambda t}\int_0^{\lambda t}\frac{1-e^{-z}}zdz
\end{equation}\label{eq:VarMean}
where
\[
\sigma^2_1(Q(t|\tau)) = \int_0^tQ^2(t|\tau)\rho_1(\tau)d\tau -\left(\int_0^tQ(t|\tau)\rho_1(\tau)d\tau\right)^2.
\]
For the sum, the result is an example of the law of total variance

\[
\sigma^2\left(\sum_{i=1}^{N(t)}Q(t|t_i)\right) = \mathbb{E}\left(\sigma^2_N\left(\sum_{i=1}^{N}Q(t|t_i)\right)\right) + \sigma^2\left(\mathbb{E}_N\left(\sum_{i=1}^{N}Q(t|t_i)\right)\right) 
\]
where the outer expectation and variance are with respect to $N(t)$ being
arbitrary, and the inner terms are conditional on $N(t)=n$.  
Thus,
\[
\sigma^2\left(\sum_{i=1}^{N(t)}Q(t|t_i)  \right) = \sum_{n=0}^{\infty}p_n(t)\left(\mathbb{E}_n\left(\sum_{i=1}^nQ(t|t_i)-n\mu\right)^2   + \left(n - (\lambda t)\right)^2\mu^2\right)
\]
where $\mu = \mathbb{E}_1[Q(t|\tau)]$. After simplification, this yields, 
\begin{align}
\begin{aligned}
\sigma^2\left(\sum_{i=1}^{N(t)}Q(t|t_i) \right) &=(\lambda t)\mathbb{E}_1[Q(t|\tau)^2]
\end{aligned}
\end{align}\label{eq:VarSum}
Thus, for a Poisson process, the mean and variance can be computed in a
straightforward manner even when $Q$ is very complicated.

\subsection{Composite random processes}
\label{subsec:ComposRand}

In the foregoing the secondary variable $x(t|t_i)$ is deterministic when $t_i$
is fixed, but this can be generalized as follows.  If the analog of the position
is a Bernoulli random variable $R(t|\tau)$,  for example $R=1$ if a cytoneme has
returned to $x_0$ and 0 otherwise, then the mean and variance if $R$ are
equivalent. In this case, it is also possible to compute the distribution
function,
\begin{equation}
S(N_r,t)\equiv \mathbb{P}\left[\sum_{i=1}^{N(t)}R(t|t_i)-N_r\geq 0  \right] = \sum_{n=0}^{\infty} p_n(t)\mathbb{P}_n\left[\sum_{i=1}^nR(t|t_i)>N_r\right]\label{eq:BernoulliDist}
\end{equation}
where $\mathbb{P}_n$ indicates a conditional probability given $N(t)=n$. We
define $\bar{r}(t)=\mathbb{E}_1[(r(t|\cdot)]$ as the probability for an
individual cytoneme to have returned  by time $t$ when it could have started at
any $\tau\in(0,t)$. The computation of the resulting distribution has been
derived in  \cite{ross1996stochastic} and an alternate derivation is presented
in the appendix. The result is that the overall expectation is exponentially
distributed with 
\[
S(N_r,t) =  1 - e^{-\lambda t \bar{r}(t)}\sum_{j=0}^{N_r-1}\frac{(\lambda t \bar{r}(t))^j}{j!}
\]
Furthermore, if $N_r$ is a fixed number, then by writing 
\[\mathbb{P}\left[\sum_{i=1}^{N(t)}R_i  = N_r  \right] =  S(N_r,t)-S(N_r+1,t) = \frac{1}{N_r!}e^{-\lambda t\bar{r}(t)}(\lambda t \bar{r}(t))^{N_r},
\]
we see that the resulting distribution for the number of packets received is
Poisson-distributed with parameter $\lambda t\bar{r}(t)$, which is the product
of the probability of a cytoneme returning with morphogen, times the expected
number of cytonemes generated in $(0,t)$. Thus, the mean and variance of the
number of packets received are $\lambda t \bar{r}(t)$ which depends on the FPT
for a cytoneme to return to the cell with a packet of morphogen. Note also that
as $t$ increases, $\bar{r}(t)$ increases and eventually approaches a number
$0<\alpha<1$ which is the probability of a cytoneme returning to a cell with
morphogen at any $t$. Thus, in the long term, the number of cytonemes that bring
morphogen to a cell approaches $\alpha \lambda t$ with $\lambda t$ representing
cytoneme generation, and $\alpha$ an efficiency factor that describes how likely
they are to find their target.

These results are also particularly interesting since the dynamics of the
cytoneme extension, adhesion, and retraction processes are all summarized in
$r(t|\tau)$ (even if $r(t|\tau)$ is for instance the complicated FPT from the
previous section), but regardless of how complex $r(t|\tau)$ is, if the cytoneme 
generation is a Poisson process, then the morphogen received will be exponentially
distributed with a parameter that depends on the cytoneme dynamics.

So far, we have assumed that variations in the amount of morphogen per cytoneme are not
important. But, if there is a significant variation, a similar approach can be
used, although a closed form is not possible in general as some of the
simplifications used to obtain the results above no longer apply. If we wish to compute the distribution,  
\[
P(m,t) = \mathbb{P}\left[\sum_{i=1}^{N(t)}q(t|t_i)\geq m\right]
\]
where $M(t|t_i)$ is the amount of morphogen returned to the cell by the cytoneme that was originated at $t_i$, then assuming that the amount of morphogen carried per cytoneme is independent of time and of any previous cytonemes, we obtain
\[
P(m,t) = \sum_{n=0}^{\infty}p_n(t)\left[ \sum_{j=0}^n  \binom{n}{j}\bar{r}(t)^j(1-\bar{r}(t))^{n-j}\mathbb{P}\left[\sum_{i=1}^j M_j \geq m |j \right]\right]
\]
where the conditional probability 
\[
\mathbb{P}\left[\sum_{i=1}^j M_i \geq m |j \right] = \int_m^{\infty} \mathcal{M}^{(j)}(m)dm
\]
is conditioned on $j$-cytonemes returning to the cell by time $t$. The symbol
$\mathcal{M}^{(j)}$ is the $j$-fold convolution of the distribution function for $M_i$.  It does not appear that this sum can be simplified to obtain a closed-form for the distribution. On the other hand, we expect that cells only produce limited numbers
of cytonemes, thus $n$ need not be taken very large and it may be possible to
numerically approximate the summation under biologically reasonable assumptions.
We did not investigate this here since the distribution of morphogen per
cytoneme is currently unknown, but we hope that this example displays the
usefullness of the stochastic analysis, and may be soon compared with
experimental results as they arise.

Finally, to conclude this section, we also mention that Equation \eqref{eq:Mean}
is particularly important since it connects the stochastic and continuum models
by giving a definition for the average number $n(x,t)dx$ of cytoneme tips in a
small interval, $(x,x+dx)$. Since the cytoneme velocity is fixed, this quantity
is equal to the number of cytonemes generated in $t\in(x/v,(x+dx)/v)$, which is
on average $\lambda dt H(t-x/v)$, as was found above.

If we allow $n(x,t)$ to be indexed as well by the starting position, $x_0$ of
the cytoneme, then $n(x,t|x_0)$ is essentially $p_e(x,t|x_0)$ from the model
given above. Let us then consider $q(t|x_0)$, and compute for a given $x_0$ the
mean and variance of this quantity. Given the continuum model above, the mean is
exactly that obtained from the continuum model. In particular, let us define
$m(t|x_0,t_0)$ as the distribution of arrival times for a cytoneme starting from
$x_0$ at time $t_0$, then
\[
q(t|x_0) = \int_0^t m(t|x_0,\tau)\lambda(\tau)d\tau = \gamma v \lambda_0 H\left(t-\frac{2x_0}{v}  \right)\left[t-\frac{2x_0}{v}+\frac{1}{\mu_d}\left( 1-e^{-\mu_d\left(t-\frac{2x_0}{v}\right)}\right)   \right]
\]
For the variance, we then have
\begin{align*}
\begin{aligned}
\sigma^2(m(t|x_0)) &= \int_0^t m^2(t|x_0,\tau)\lambda(\tau)d\tau \\
&= \int_{2x_0/v}^t\left[ \frac{\gamma v}{\mu_d}\left(1-e^{-\mu_d\left(t-\frac{2x_0}{v}\right)}\right)\right]^2\lambda_0 d\tau \\
&=  \frac{\lambda}{\mu_d}\left(tv-2x_0-\frac{v}{\mu_d}\left(3+e^{-2\mu_dt+\frac{4\mu_d x_0}{v}}-4e^{-\mu_d t+\frac{2\mu_d x_0}{v}} \right)\right)H\left(t-\frac{2x_0}{v}\right)
\end{aligned}
\end{align*}
%


A similar analysis is valid if we allow a spatial Poisson  process
to describe generation between multiple cells at various starting points,
$x_0$. In that case, we introduce a spatial number density $\mu^{(1)}(\bsm{x})$
which is defined such that for any set $B\subset \mathbb R^n$, we have 
\[
\mathbb{E}\left[N(B)\right] =\int_B \mu^{(1)}(\bsm{x})d\bsm{x}. 
\]
We can also define similar conditional densities on subsets, $B$,
e.g. $\nu_B^{(1)} = \mu^{(1)}(x)/\int_B\mu^{(1)}d\bsm{x}$. Let us assume that
the spatial and temporal components are decoupled, e.g. the density
$\rho^{(n)}((x_1,t_1),\dots,(x_n,t_n)) =
\rho^{(n)}(t_1,\dots,t_n)\nu^{(n)}(x_1,\dots,x_n) = \prod_{i=1}^n
\rho^{(1)}(t_i)\nu^{(1)}(x_i)$. Likewise, we obtain the mean,
\[
\mathbb{E}\left[\sum_{i=1}^n r(t_i,x_i)  \right] = \lambda\mu\int_{\Omega}\int_0^t r(\tau,x)d\tau\ dx
\]
which is analogous to the mean computed above, except that we must now integrate
over space as well as time. The variance follows this same trend.  While
statistics of this compound stochastic process can be computed analytically in
these simple cases where all of the processes are independent of one another, it
is clearly of scientific interest to incorporate interactions between competing
processes as well. While perturbation expansions of the above results may be
feasible for weak interactions, strong interactions require an alternative
approach since it is not valid \emph{a priori} to assume that strong
interactions can be modeled via a series expansion with respect to a
non-interacting system. Thus, we next develop a
stochastic simulation algorithm that can deal with general interactions.

\subsection{Stochastic simulations }                                                                                                                                                                                                                                                                                                                                                                                                                                    
\begin{figure}
 \begin{center}
\includegraphics[width=.4\textwidth]{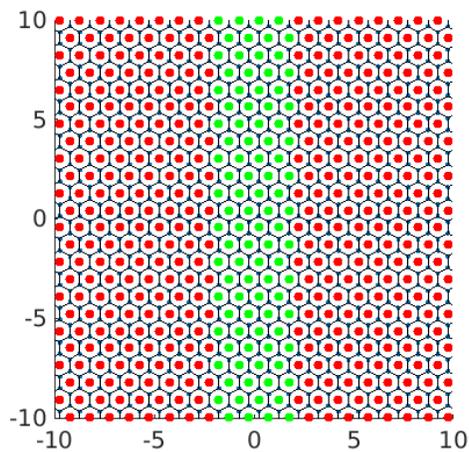}
\end{center}
  \caption{The underlying grid for the problem. Cells marked in green are producer
    cells, all other cells are receivers.} 
  \label{grid}
\end{figure}

As shown above, analytical results can be obtained when the cytoneme models obey certain assumptions so that the equations are workable.  
  However, with an eye towards more
 complex models, for instance, where there are interactions between cytonemes,
 or when there is additional signalling occurring, we developed a stochastic
 simulation algorithm that considers the dynamics and internal states of each
 individual cytoneme to study cytoneme-based morphogen transport. We first
 consider the RIT problem, and to begin we lay down a hexagonal array of cells,
 with a stripe of morphogen producing cells located near the $y$-axis in the
 center of the domain and receiving cells located on either side of this
 strip (see Figure \ref{grid}).

\begin{figure}
\begin{center}
\includegraphics[width = 0.55\textwidth]{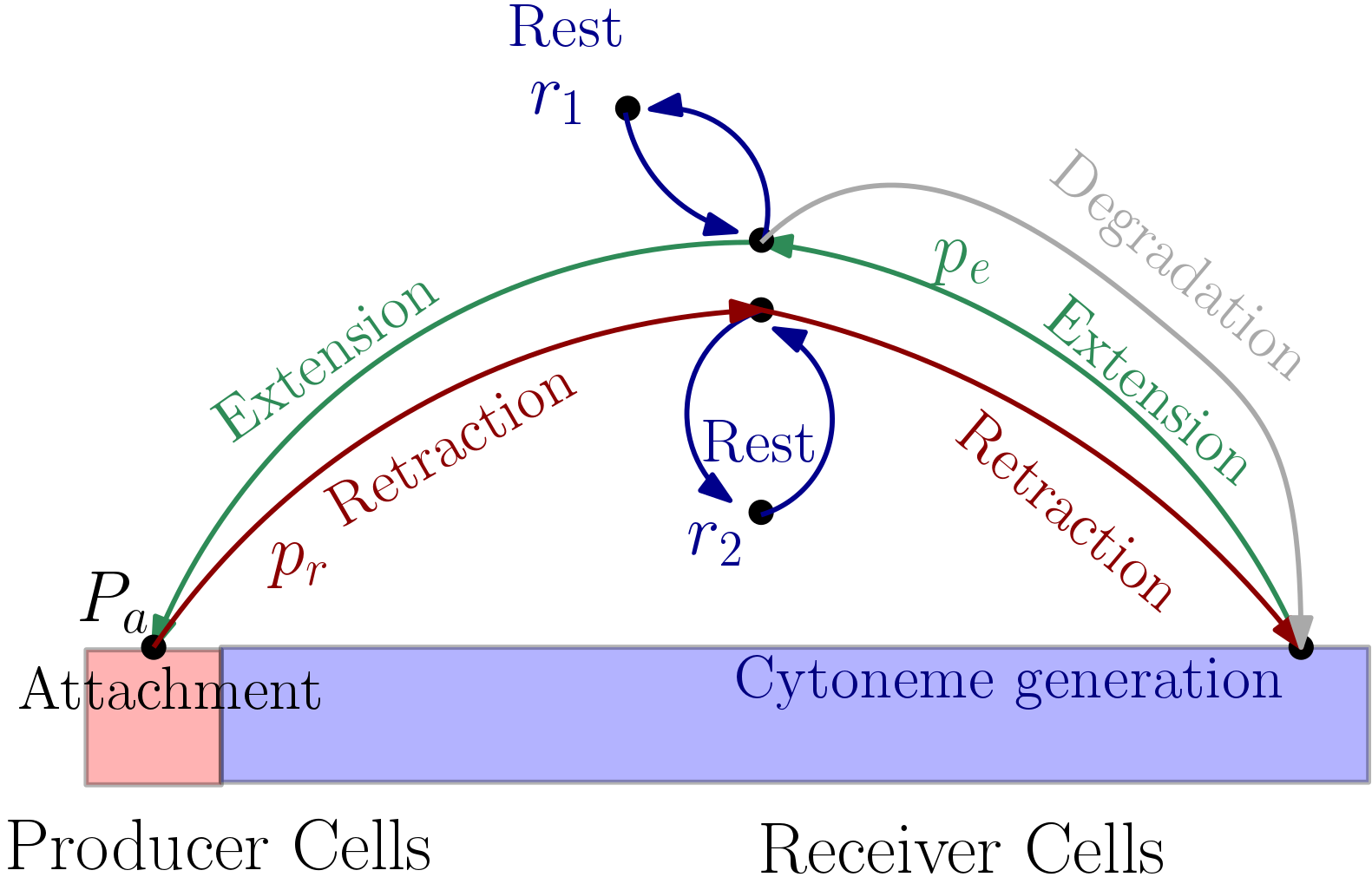}
\end{center}
\caption{A diagram of the various transitions that can occur in the RIT
  model. Cytonemes start in the extension phase after generation. At any point,
  they may enter a rest state at  a rate $r_1$, or spontaneously begin retracting without
  reaching the source (degradation). If they reach the source, they become
  attached, and then undergo retraction, or enter  intermittent rest phases
  at a  rate $r_2$ as they retract.   \label{fig:RIT_Transitions}} 
\end{figure}

The following algorithm is used to update the temporal evolution of the cytoneme
generation and cytoneme position processes in   increments of
length $\delta t$.  The model parameters and typical values in dimensionless
form are given in Table \ref{tab:StochParams}, and a diagram of the possible
state changes is shown in Figure \ref{fig:RIT_Transitions}.
\noindent\fbox{%
    \parbox{\textwidth}{%
\begin{enumerate}
\item Each receiver cell generates a new cytoneme with probability $p_i = 1-e^{-\lambda \delta t}$
\item For all cells that generated a new cytoneme, the directional orientation of that cytoneme is chosen from a uniform distribution $\mathcal U(-\theta_0,\theta_0)$ so that the
cytonemes can emanate from any cell in a cone-shaped region of angular width $2\theta_0$ opening towards the
source cells.
\item For each cytoneme, the position is updated at each time step as
$\bsm{x}(t+\delta t) = \bsm{x} - \delta t \bsm{v}(\ell,\theta,t)$ where $\ell =
\vert\bsm{x}(t)-\bsm{x}(t_0)\vert$ and $t_0$ is the time the cytoneme was
initiated. The velocity could simply be a constant, or could be a function of direction, length and time. 
\item For each cytoneme, there is also a probability  $p_f = 1-e^{-\lambda_r \delta t}$ that the
cytoneme will go from extension to retraction at each time step. 
\item When a cytoneme is located near a producing cell, there is an additional
probability $p_a = 1-e^{-\mu \delta t}$ of attaching to a producing cell
and receiving a morphogen-packet. We assume that all morphogen packets carry
equal amounts of morphogen at their origin,  and thus   the total amount 
each receiving cell gets is  proportional to the number of packets it has 
received.
\item If a cytoneme is successful at obtaining a packet, then it
switches to the retraction state, and its position is updated at each time step
by $\bsm{x}(t+\delta) = \bsm{x}(t)-\bsm{v}_r\delta t$
\item If a cytoneme length reaches $\ell=0$, that cytoneme is eliminated. If that cytoneme was carrying a morphogen, one morphogen packet is added to the total received by the cell that originated the cytoneme.
\end{enumerate}
}
}

\begin{table}[h!]
\begin{center}
\begin{tabular}{|c|c|c|}
\hline
Symbol & Definition & Default Value\\
\hline
$\lambda$ & Cytoneme generation rate & $0.4 $ \\
\hline
$\mu$ & Cytoneme attachment rate & $100$ \\
\hline
$v_0$ & cytoneme velocity & $1$ \\
\hline 
$\lambda_r$ & cytoneme halting rate & $1$ \\
\hline
$\theta$ & cytoneme angle & $-\pi/8\leq \theta \leq \pi/8$\\
\hline
$\delta t$ & time-step length & $0.01$\\
\hline
\end{tabular}
\end{center}
\caption{Typical parameter values for the stochastic simulation
  method. Typical length scales are on the order of a $\mu m$, and time scales on the order of 1 second.
\label{tab:StochParams}}
\end{table}

This simple stochastic simulation approach is straightforward to implement and
allows for simulations with thousands of producing and receiving cells that
approach realistic scales that can be compared with biological tissues. A key
result from a  simulation is that we can in a straighforward manner obtain the
distribution of morphogen, and of cytoneme lengths and positions throughout the
tissue dynamically as the simulation evolves.

\begin{figure}
\begin{center}
\includegraphics[width = 0.5\textwidth]{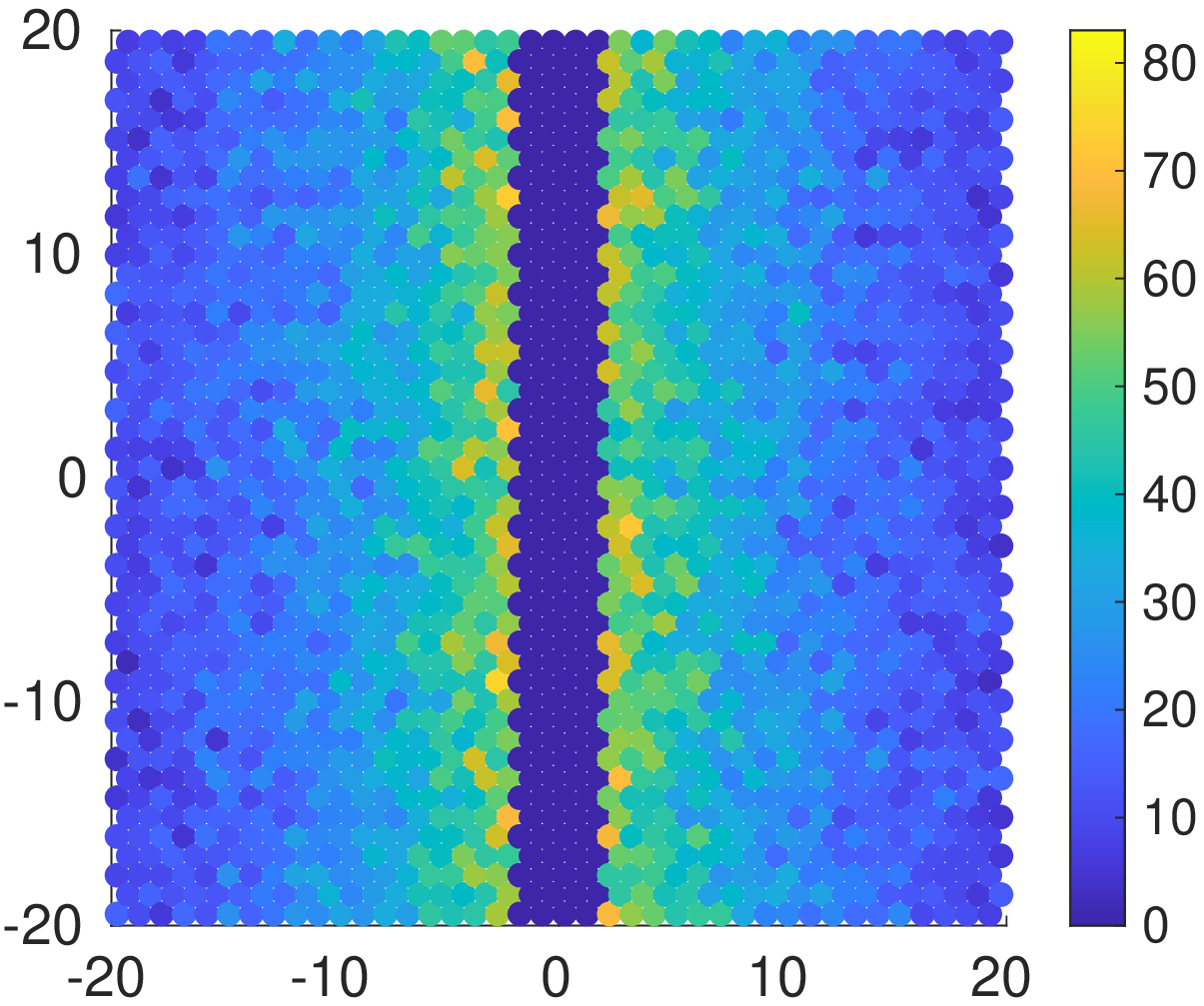}\\
 \includegraphics[width  = 0.45\textwidth]{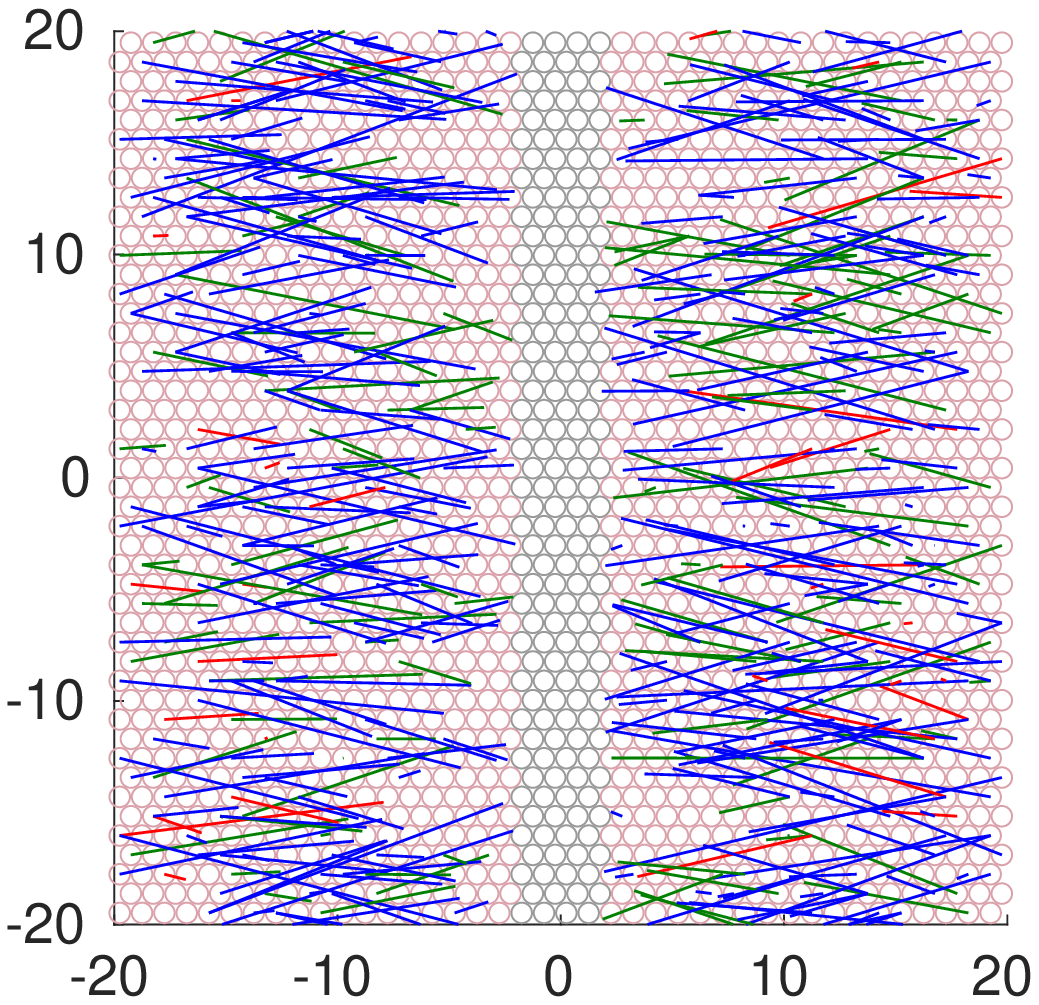}
\end{center}
\caption{ Depiction of the result of a typical simulation with 920
  cells. (Top) Coloring
  corresponds to the number of morphogen packets received and the colorbar gives
  the number scale. The center column of cells  are
  the source cells. (Bottom) Diagram of the cytonemes during a simulation at t =|rcol{xx}.. Black
  cytonemes are extending, red are retracting without morphogen, and green
  received a morphogen packet and are retracting.
\label{fig:MDist2D}}
\end{figure}

The morphogen distribution from a typical simulation is shown in Figure
\ref{fig:MDist2D} to illustrate the typical output. At the top the  source
strip is the column of cells at the center and outside that each dot
represents  a single cell, colored according to the number of 
morphogen packets received.  It  is clear that the
amount of morphogen received decays with distance from the source region,

The spatial distribution of the cytonemes is  shown  in the bottom of Figure
\ref{fig:MDist2D}. In particular, cells close to the source can rapidly extend
cytonemes, receive morphogen and retract, thus it appears at any point in time
that there are less cytonemes emanating from the closest cells, even though they
receive the most morphogen overall. Additionally, we see that more of the red
retracting cytonemes (those that never reached a source cell) emanate from more
distant cells, indicating that the likelihood for a successful morphogen transfer
depends on distance from the producing cells,  as expected.

\begin{figure}[h!]
\begin{center}
\includegraphics[width = 0.3\textwidth]{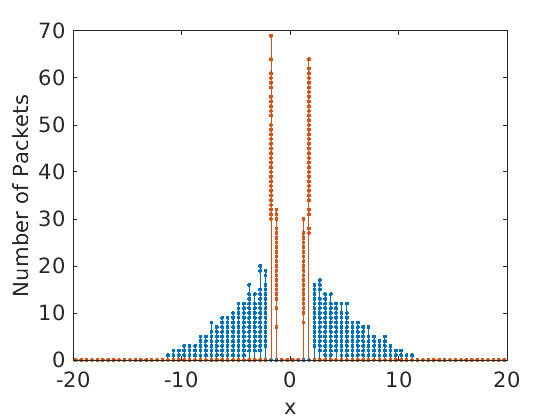}
\includegraphics[width = 0.3\textwidth]{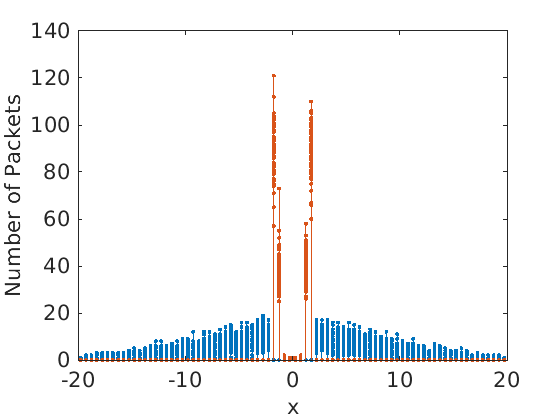}
\includegraphics[width = 0.3\textwidth]{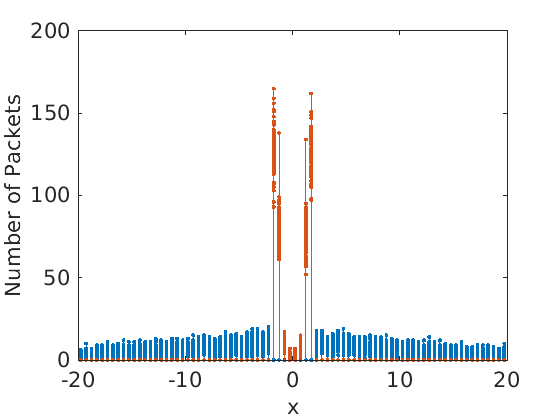}
\\ \includegraphics[width = 0.3\textwidth]{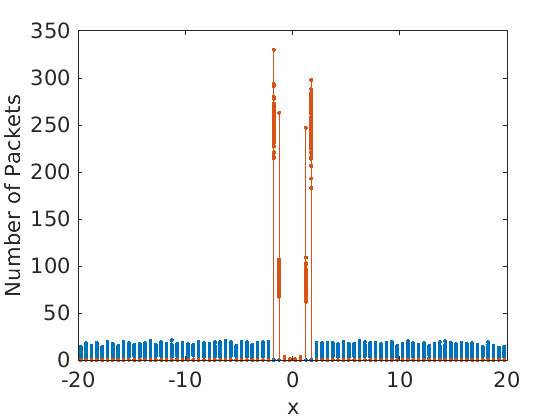}
\includegraphics[width = 0.3\textwidth]{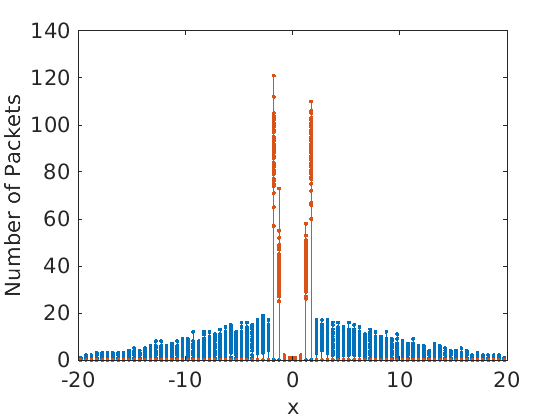}
\includegraphics[width = 0.3\textwidth]{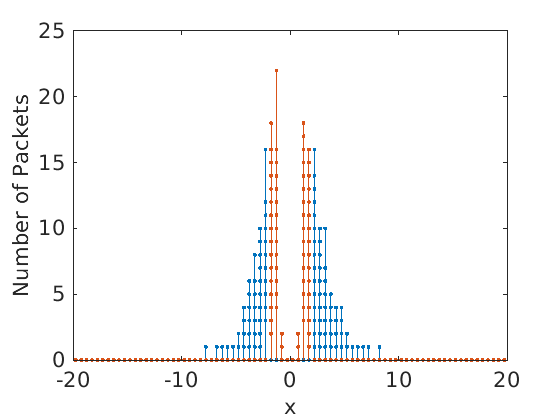}
\\ \includegraphics[width = 0.3\textwidth]{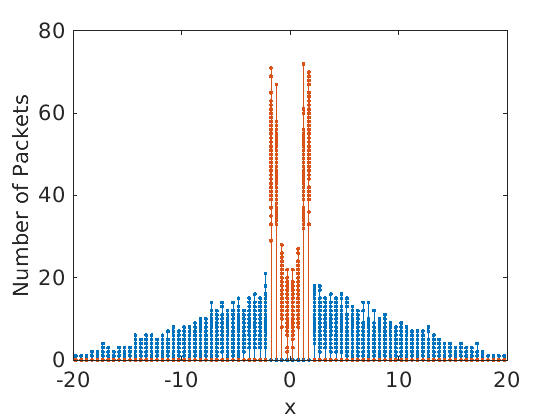}
\includegraphics[width = 0.3\textwidth]{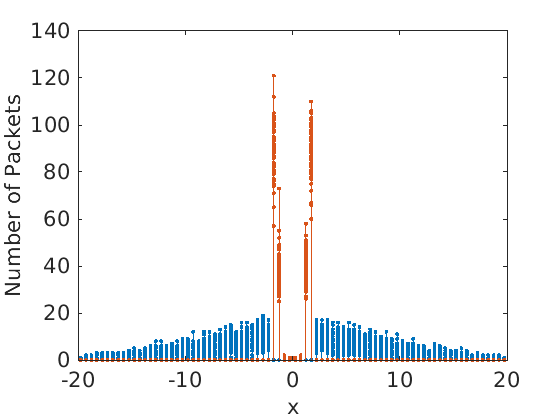}
\includegraphics[width = 0.3\textwidth]{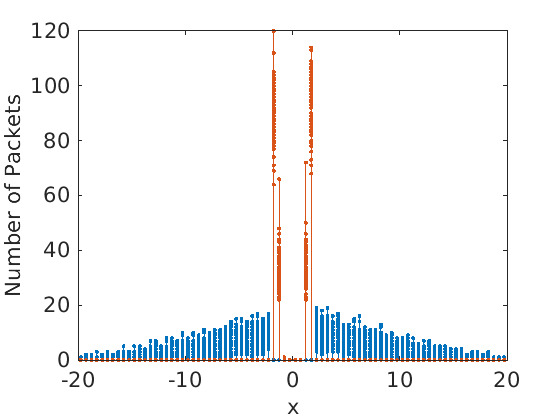} \\
\end{center}
\caption{The  spatial distribution of morphogen as a function of distance from the
  morphogen source under several different combinations of the parameters. In
  each row from right-to-left, one parameter is increasing. The top row has the
  initial cytoneme velocity, $v_0\in\{0.5,1,2\}$. The
  middle row has the stopping rate for cytonemes in the range $\lambda_r =
  \{0.01,0.1,1\}$, and the bottom row has the attachment rate in the
  source region in the range  $\mu = \{1,5,10 \}$
  \label{fig:MDist}}
\end{figure}

Collapsing the morphogen distributions in Figure\ref{fig:MDist2D}  to a one-dimensional
distribution for each value of $x$ leads to the  distributions
shown in Figure \ref{fig:MDist} for different parameters governing the
morphogen generation rate, velocity parameters, and likelihood of receiving a
morphogen packet.
The main parameters that are varied there  are the starting velocity of
the cytoneme ($v_0$),  the cytoneme halting rate ($\lambda_r$),
and the attachment rate in the source region ($\mu$). The following
observations emerge from this.
Increasing $v_0$
tends to yield flatter profiles of longer extent. Increasing $\lambda_r$
leads to shorter morphogen distribution profiles, and much less morphogen
transferred per unit time, and increasing $\mu$ only changes the distribution of
the number of packets sent from cells within the source region. In particular,
with high $\mu$, only the source cells at the boundary of the source region
will make many attachments, since the attachment dynamics are fast enough to
make it unlikely that any cytonemes pass over the boundary to the interior. In
most cases, we also see that the variance of the morphogen packets received
scales with the mean number of morphogen packets received for cells located at
position $x$.

\begin{figure}
\begin{center}
\includegraphics[width = 0.3\textwidth]{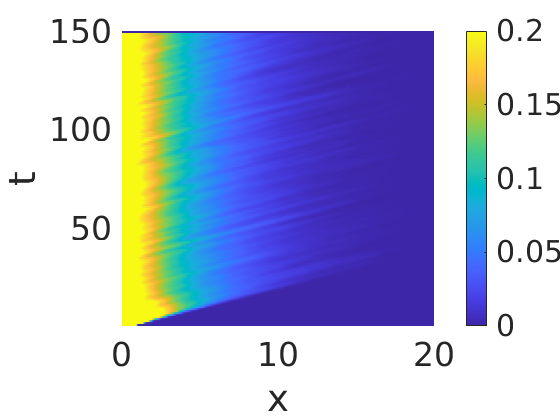}
\includegraphics[width = 0.3\textwidth]{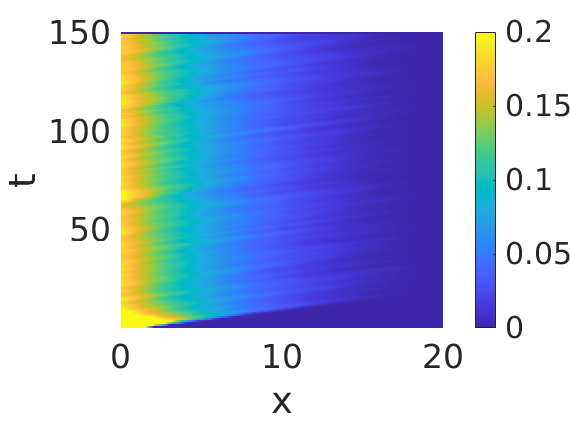}
\includegraphics[width = 0.3\textwidth]{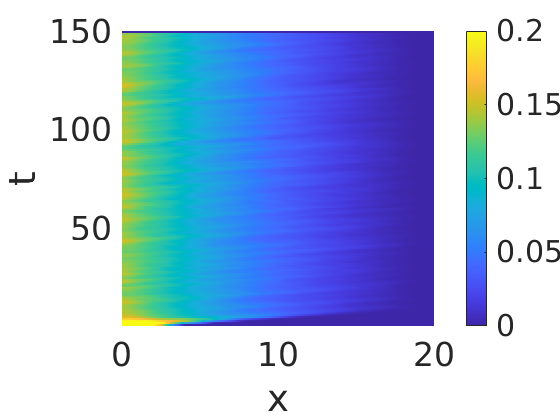}
\end{center}
\caption{The cytoneme length distributions as functions of $\ell$ and $t$. On
  the left $v_0 = 0.5$ and $\mu_0 = 0.1$, in the middle $v_0 = 1$ and on the
  right $v_0 = 2$. \label{fig:LengthDist}}
\end{figure}

Aside from the distribution of morphogen, it is also interesting to study the
length distribution of cytonemes in the tissue. Understanding the length
distribution may serve as a means to distinguish different transport processes
that may yield similar morphogen profiles - for instance whether the transport
is receiver- or producer-initiated. The cytoneme length distribution is obtained
via a non-parametric kernel smoothing approach. The length of every currently
existing cytoneme is computed at each step of the simulation,  and we approximate
the length distribution for that step as
\[
p_b(t,\ell) = \sum_{k=0}^{N_c}k_b(\ell-\ell_k)
\]
where $b$ is a bandwidth parameter that governs how smooth the resulting
approximate distribution is. In practice, the dynamics are quite noisy if this
is simply computed at each time step. Thus, we also apply smoothing in the
time-domain by summing over $t$ for a range of values, e.g.
\[
p_{b,m}(t,\ell) = \sum_{s=t-m\delta t}^{s = t+m\delta
  t}\sum_{k=0}^{N_c(s)}k_b(\ell-\ell_k)
\]
The resulting distributions are shown as a function of time in Figure \ref{fig:LengthDist}
for several different parameter ranges.
Notice that as the velocity increases (from left to right),
  the distribution becomes less concentrated near $\ell=0$, and hence a broader morphogen
  distribution, as expected.  We also see that with a faster velocity, the 'steady-state'
  length distribution is  achieved more rapidly.
  
\subsection{Cell saturation effects}
In the above results, the cytoneme generation dynamics and the cell morphogen accumulation
are independent. For receiving cells, this means that
the amount of morphogen received has no bearing on the cells rate of cytoneme
generation, and for source cells, the number of packets transferred has no
effect on their ability to transfer more packets as subsequent cytonemes
attach. In reality, there can be limiting steps in each case.

\begin{figure}[h!]
\includegraphics[width = 0.3\textwidth]{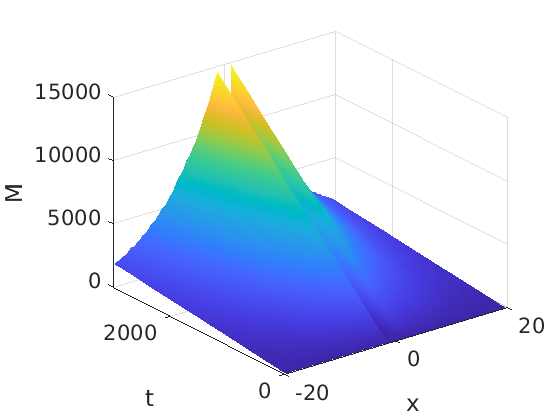}
\includegraphics[width = 0.3\textwidth]{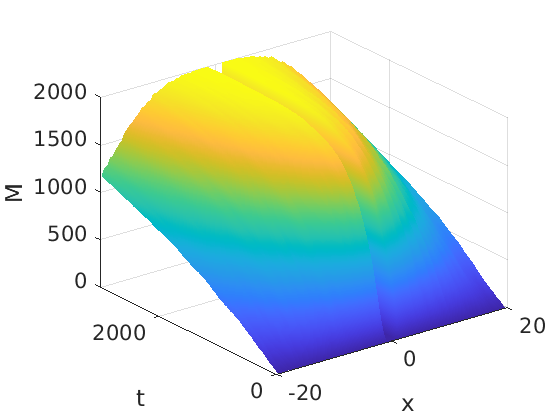}
\includegraphics[width = 0.3\textwidth]{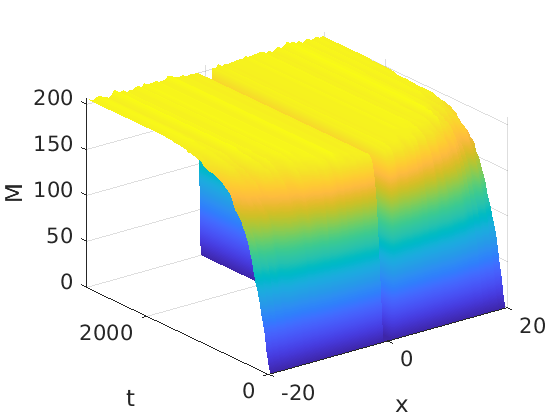}
\caption{Depiction of the space-time dependence of the morphogen
  distribution. With increasing negative feedback, one can see that the
  concentration rapidly levels off. Feedback is increasing from left to right.\label{fig:Feedback}}
\end{figure}

Negative feedback loops are commonly found in morphogenetic systems. For
receiving cells, this often leads to a control system where, as morphogen is
received, the cell down-regulates the receptors for that morphogen. This could
lead to saturation effects, and in a biological setting may have the purpose of
modifying the morphogen distribution such that more distant cells can receive
morphogen rather than all of it being absorbed by cells near the source. In our
simulations, this effect can be implemented by modifying the cytoneme generation
rate of cells as a function of the amount of morphogen they have received.  This
effect leads to a traveling front, where row-by-row, cells reach their
saturation point leading to an expanding region of cells at nearly constant
morphogen concentration. This is depicted in Figure \ref{fig:Feedback}, and
described in more detail below.

When cells can shut-off their cytoneme machinery after receiving a certain level
of morphogen, there is a significant dynamic behavior that differs from the
previous case. Rather than attaining an equilibrium length distribution, and a
morphogen distribution that simply scales with time, in this case there is a
traveling-wave like propogation of cell-saturation. This starts with the closest
receiver cells to the producer region which attain saturation morphogen levels
rapidly followed by the gradual saturation of increasingly distant
cells. Finally, the cytoneme production turns off once nearly all cells have
achieved saturation.

To highlight this effect, the length and morphogen distribution are depicted as
functions of time and length (position) showing how the evolution varies in 1)
no saturation, 2) high morphogen saturation, and 3) low saturation. In the
no-saturation case, the result is as before. In the high-saturation case, a
large amount of morphogen is required to saturate a cell, so the results also
appear fairly similar for a long time until eventual saturation is reached. In
the low saturation case, only a small number of morphogen packets are needed to
saturate a cell, and the behavior is distinctly different.

It is also interesting to consider what effect morphogen production dynamics
play in the producing cells.  The most direct approach appears to consider that morphogen packets are produced according to a stochastic process that has
a saturation point, e.g.
\[
\frac{d}{dt}P(M=m,t) = \lambda(m-1)P(m-1,t)-(F(t)+\lambda(m))P(m,t)+F(t)P(m+1,t)
\]
where $\lambda(m)$ is a decreasing function of $m$ that tends to zero at some
maximum concentration $M^{\ast}$. The function $F(t)$ refers to the stochastic
cytoneme attachment process - when a cytoneme attaches to a cell that has  $M+1$
packets, the number of packets decreases to $M$.  The morphogen generation
process could be simulated numerically by setting the probability of a new packet
being developed in $(t,t+\delta t)$ equal to $1-e^{-\lambda(m)t}$.
However, we did not conduct further simulations along these lines as there is little experimental evidence with which to find appropriate parameter values for $\lambda(m)$. Thus, this is an area of future interest, and for situations when many cytonemes can attach to a
producer cell, it may be necessary to consider producer cell dynamics as a
limiting factor on the rate at which morphogens are distributed. 

We will now turn to a discussion of diffusive transport mechanisms, followed by some comparisons between diffusion and cytoneme-based methods of distributing morphogens. 

\section[Transport in complex environments]{Diffusion vs cytonemes in a complex tissue environment}

Both the Turing and positional-information mechanisms are based on diffusion 
in a homogeneous medium. The latter posits a mechanistic description of the form
given below for establishment of the morphogen profile,
\begin{alignat}{3}
\label{1dunscalt}
\dfrac{\partial c}{\partial t} &= D\dfrac{\partial^2 c}{\partial x^2} - k c
\qquad & & x \in (0,L) \nonumber \\
&-D \dfrac{\partial c}{\partial x}(x) = j \qquad && x = 0 \\
&-D \dfrac{\partial c}{\partial x}(x) = 0 \qquad && x = L, \nonumber
\end{alignat}
but this rarely suffices to describe the detailed processes that are involved in
biological tissues. For example, extracellular signals are detected by
receptors, and this simple process adds terms for binding and release from the
receptor to Equation \ref{1dunscalt}, as well as a separate equation for the
binding and release at the cell.

\begin{figure}[h!]
  \centerline{\hspace*{-.75in}(A) \hspace*{1.75in} (B)\hspace*{2in} (C)}
  \centerline{
   \includegraphics[width=1.in]{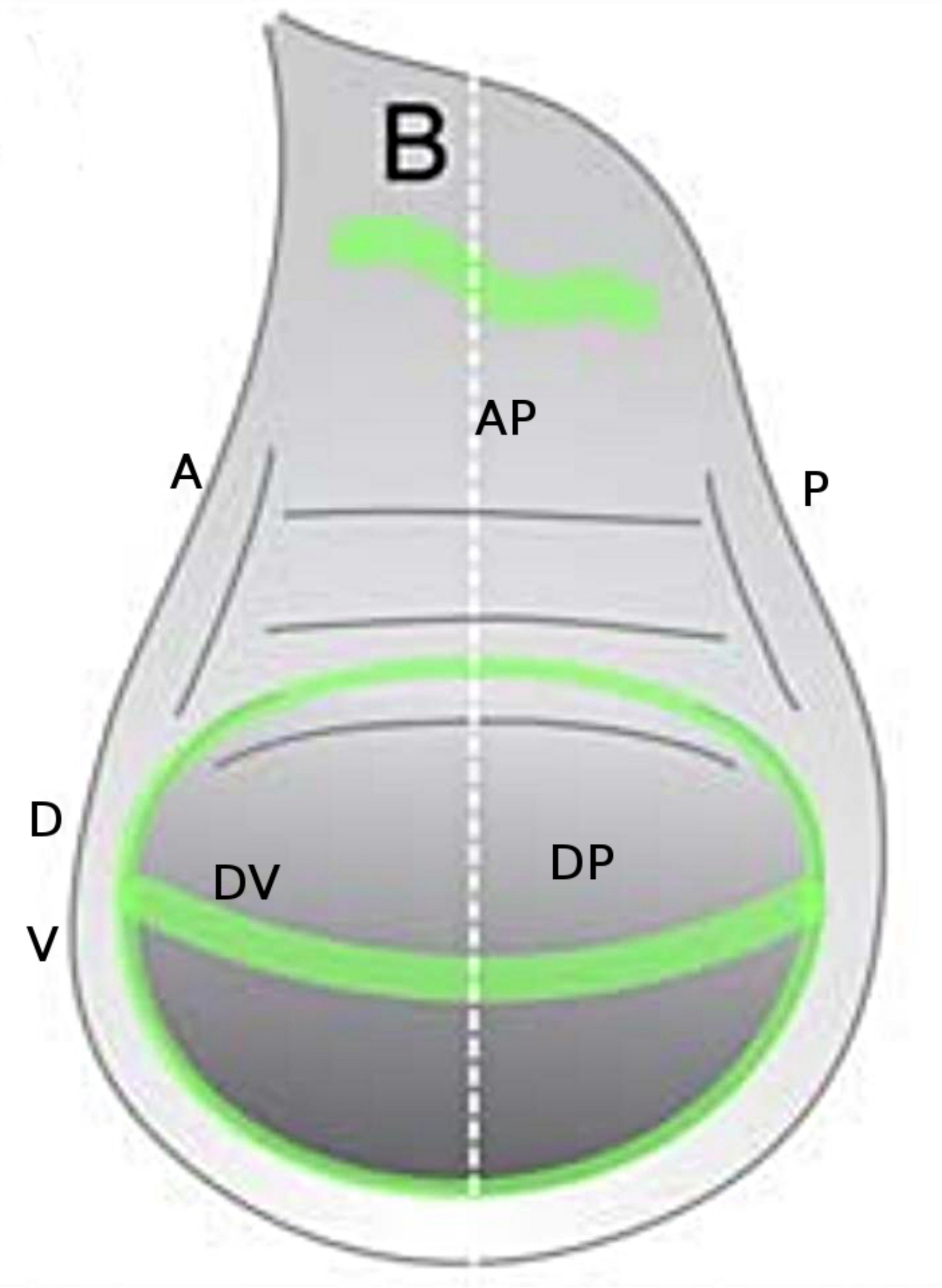}\hspace*{.25in}
   \includegraphics[width=2.25in]{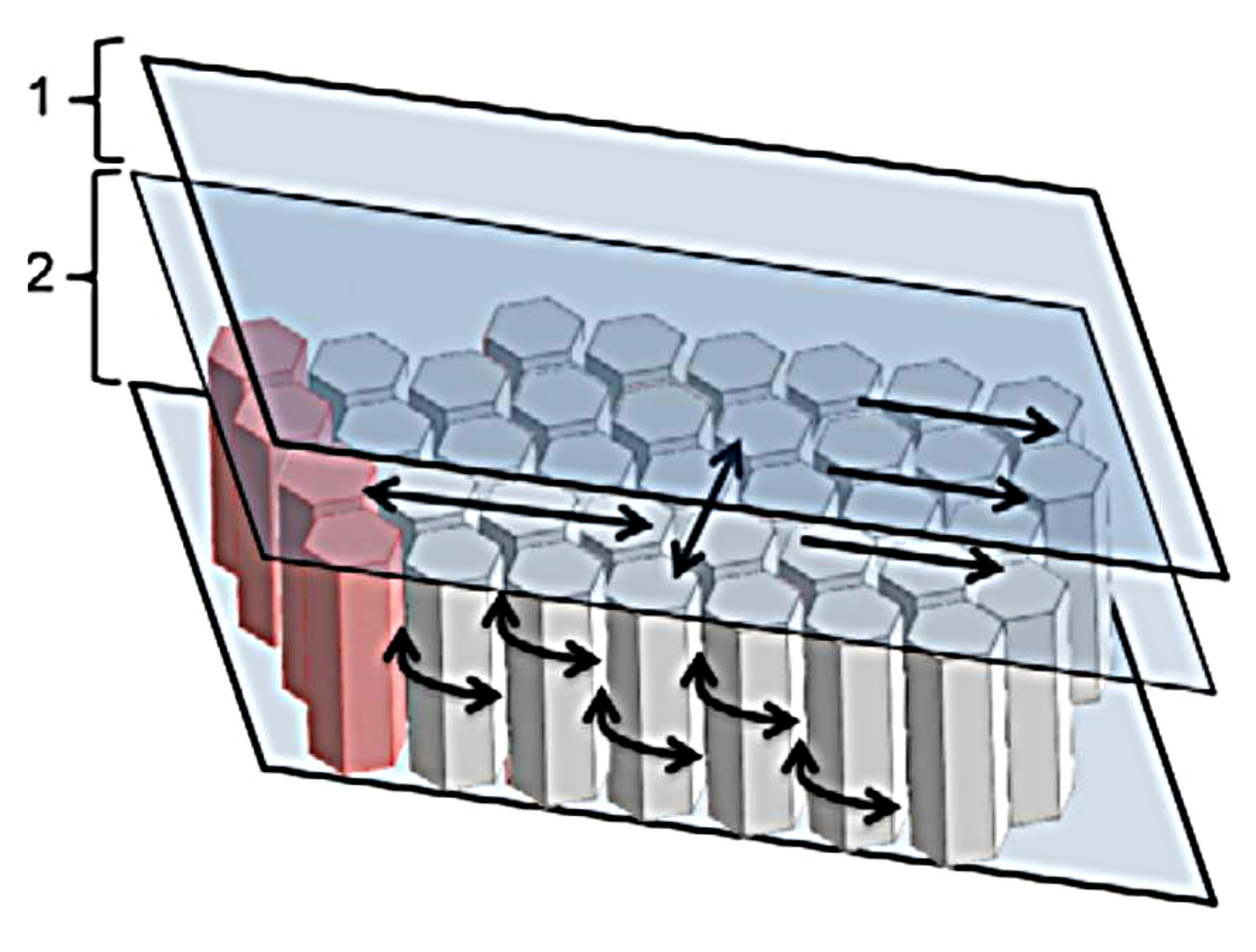}\hspace*{.25in}
   \includegraphics[width=2.25in]{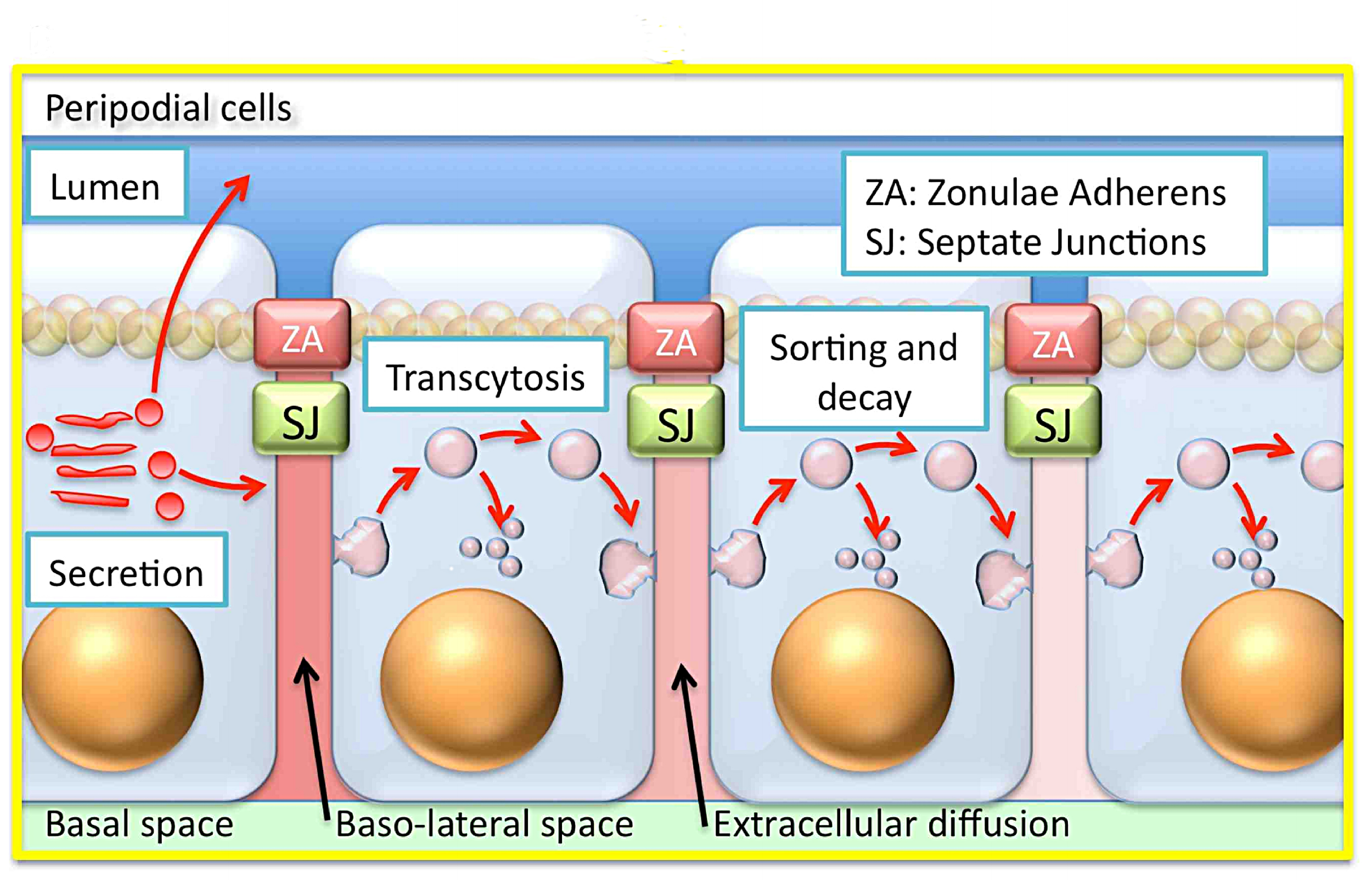}
  }
\caption{\footnotesize (A) The wing disc. (B) A section of the disc, showing the
  fluid luminal section (1), the hexagonal cells below (2), and the source of
  Dpp at the left along the AP boundary in (A). (C) A vertical cross-section
  showing the transport processes that affect the morphogen
  distribution  \cite{Othmer:2009:ITA}.  }
\label{discfig}
\end{figure}

An example of a complex tissue in which patterning occurs is the widely-studied
{\em Drosophila melanogaster} wing disc shown in Figure \ref{discfig}.  The disc
has two cell layers separated by a fluid lumen (Figure \,\ref{discfig}(A)), one
a layer of columnar epithelial cells with the apical side at the lumen and a
peripodial epithelium overlying the lumen \cite{Gibson:2002:LTD}(Figure
\,\ref{discfig}(B)). The lateral membranes of adjacent cells are connected via
two classes of junctions (ZAs and SJs) that separate the extracellular fluid
into apical and baso-lateral layers
\cite{Gibson:2009:CTG,Harris:2010:AJM,Choi:2018:UPH} (details of the entire
developmental process are given in \cite{Gou:2020:GCD}).

The wing disc is perhaps the best understood system in which transport of a
molecule -- in this case the morphogen Dpp, involves a number of very distinct
steps in a geometrically-complex tissue.  The secretion of Dpp from a stripe of
cells along the AP boundary, colored red in Figure \,\ref{discfig}(b) gives rise to
a spatial distribution of Dpp transverse to the anterior-posterior midline  of the disc, but how the
distribution is established is still an open question.  Three modes of transport
have been identified in the disc -- diffusion in the extracellular space,
transcytosis, and transport via
cytonemes \cite{kicheva2007kinetics,Schwank:2011:FLR,Zhou:2012:FED}.  While
cytonemes are important in the transport of other
morphogens \cite{Bischoff:2013:CRE}, their role in Dpp transport is more
ambiguous \cite{Akiyama:2015:MTT,matsuda2016bmp}.  Diffusion in the extracellular
space can be free Brownian motion of a particle in solution, called 'free
diffusion', or it may involve interactions with other factors, which is called
'facilitated diffusion' or 'restricted
diffusion' \cite{Shimmi:2005:FTD,Schwank:2011:FLR,restrepo2014coordination}.
 
Some experimental data shows a gradient of Dpp in the lumen and an apical layer
of the columnar cells, but other results show that a graded Dpp distribution is
found in the baso-lateral (BL) space
\cite{mundt2013characterization,Harmansa:2017:NBT} and the lumenal Dpp is
uniformly distributed \cite{Gibson:2002:LTD,Harmansa:2017:NBT}.  Still others
have shown that the majority of Dpp is in the intracellular space and the
intracellular apical Dpp forms a long-range gradient
\cite{entchev2000gradient,kicheva2007kinetics}, but some have observed the
opposite \cite{belenkaya2004drosophila,Muller:2013:MT,Stapornwongkul:2021:GEM}.
Experimentally-measured profiles of Dpp are usually described with a
reaction-diffusion model such as Equation \ref{1dunscalt}
\cite{kicheva2007kinetics,Wartlick:2011:UMG}, but the estimate of the diffusion
coefficient needed to fit the data is is much lower then usual values for free
diffusion.  Others \cite{Zhou:2012:FED} have measured a larger free coefficient
for free diffusion in solution, and showed that the low value could be explained
if receptor-mediated uptake and degradation are incorporated. However their
analysis assumes a simple geometry corresponding to the luminal surface, but as
indicated earlier, the geometry is quite complicated.  Our next objective is to
show how the structure of a complex tissue is reflected in the macroscopic
parameters that appear in Equation \ref{1dunscalt}.

\subsection[Microscale to Macroscale]{From micro- to macro-parameters for diffusion and reaction}

 In a previous paper \cite{Stotsky:2021:RWA}, hereafter referred to as \mI, we
 analyzed how microscale processes that are involved in transport in complex
 tissues are reflected in the macroscopic diffusion and decay terms in an
 equation such as Equation \ref{1dunscalt}.  Before providing an example to
 compare with the cytoneme transport, we briefly review some of the underlying
 ideas in that work.

The cellular level models are based upon  continuous-time random walks (CTRWs).
In a CTRW model, the key quantity of interest is the spatial distribution,
$\bsm{P}(\x,t|0)$ of particles at some time $t$, given that the particles move
randomly via a series of jumps or transitions determined by a given stochastic
model.  Thus, let $\bsm{\Psi}(\bsm{x},t|\bsm{y},0)$ be the joint
probability density for a walker to instantaneously jump at $t^-$ to a point
$\bsm{x}$ after having starting at position $\bsm{y}$ at $t=0^+$. From the
space-time  distribution of $\bsm{\Psi}$ one can obtain a spatial jump distribution
$T(\bsm{x},\bsm{y})$ and the density of a waiting time distribution (WTD) $\phi(\bsm{y},t)$ as
follows.
\[
\bsm{T}(\bsm{x}|\bsm{y}) = \int_0^\infty
\bsm{\Psi}(\bsm{x},\tau|\bsm{y},0)d\tau, \ \ \ \ \phi(\bsm{y},t) =
\int_{\mathbb{R}^n}\Psi(\bsm{x},t|\bsm{y},0)d\bsm{x}.
\]
Here $\bsm{T}$ defines the probability of landing at $\bsm{x}$, having started
at $\bsm{y}$, and $\phi(\bsm{y},t)dt$  is the probability of a jump in
$(t,t+dt)$. 
From the latter one obtains the cumulative and complementary waiting-time distributions 
\begin{equation}
    \Phi(\y,t) = \int_{0}^{t} \phi(\y,s)\,ds = Pr \{ {\cal T} \le t\}
\label{PHI}
\end{equation}
and
\begin{equation}
  \hat{\Phi}(\y,t)
    = \int_{t}^{\infty} \phi(\y,s)\,ds = 1 - \Phi(t) = Pr\{{\cal T}\ge t\}.
    \label{Phihat}
    \end{equation}

In many contexts   $\phi(\bsm{y},t)$ is independent of $\bsm{y}$ and
$\bsm{T}(\bsm{x},\bsm{y})$ is only a function of $\bsm{x}-\bsm{y}$, and with these
simplifications one can obtain  the following renewal equation for the spatial
distribution of $\bsm{P}$ as
\[
\bsm{P}(\x,t|0) = \hat{\Phi}(t)\delta (\x) + \int_{0}^{t} \int_{R^n} \phi(t -
\tau)\bsm{T}(\x,\y)\bsm{P}(\y,\tau | 0) \,d\y \,d\tau.
\]
Since this equation is linear, we can define an equation for $N(\bsm{x},t)$, the
probability distribution when the initial distribution is arbitrary.  If we
suppose that the WTD is Poisson-distributed, e.g. $\phi(t)=\lambda e^{-\lambda
  t}$, one can show that the system simplifies to
\[
 \dfrac{\partial N(\x,t) }{\partial t} = -\lambda N(\x,t) + \lambda
 \int_{R^n}T(\x,\y) N(\y,t) \,d\y.
\]

\begin{figure}[h!]
\begin{center}
\includegraphics[width = 0.5\textwidth]{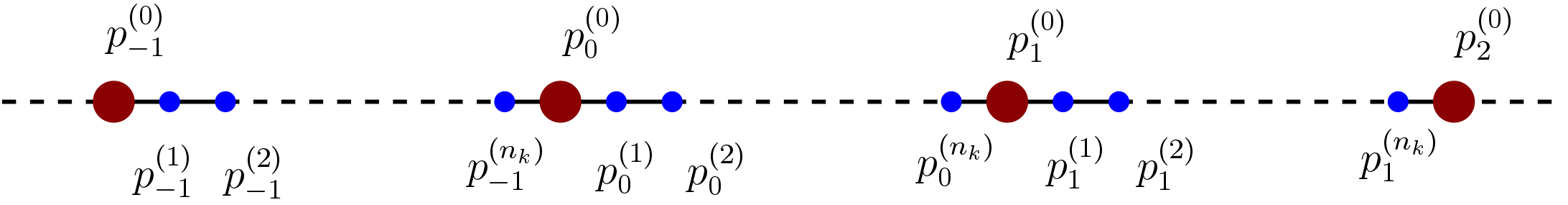}
\caption{A diagram of a 1D lattice consisting of
  primary vertices or junctions (red) and secondary vertices (blue). The
  probabilities at the primary vertices are $p_i^{(0)}$, and the secondary
  vertices are denoted by $p_i^{(j)}$.  Taken from \cite{Stotsky:2021:RWA} with
  permission. }
\label{1D lattice}
\end{center}
\end{figure}

This formulation assumes a translation-invariant continuum, but continuous time
random walks can also be posed on lattices. A lattice is a countable set of
points $\{\bsm{X}\}$ in $\mathbb{R}^d$ -- called vertices of the lattice --
whose positions are linear combinations of $d$ linearly independent vectors with
integer coefficients, endowed with a graph structure which specifies the
connectivity of these points. The term `lattice' also implies that the
connectivity is the same for each point, e.g. in a square lattice, each point is
connected to its neighboring points to the left, right, above, and below. In
\mI, \ the lattice vertices are called junctions to distinquish them from
secondary vertices that were introduced on the edges connecting junctions ({\em
  cf.}  Figure \ref{1D lattice}). Lattices are of interest biologically since in
many tissues cells arrange into lattice-like structures.

In such settings the jump kernel $T(\bsm{x}-\bsm{y})$ is defined only for the
vertices and can be thought of as a matrix operator. For a Poisson-distributed
jump rate, the evolution equation on a general lattice is of the form
\[
 \dfrac{dP_i}{d t} = \lambda\sum_{j \in {\mathcal N} (i)}T_{i.j}\left\{P_j(t) -
 P_i(t)\right\}.
\]
where $j \in {\mathcal N} (i)$ corresponds to a neighbor of $i$.  To apply this
  in biological settings, we begin with a graph or lattice structure that
  defines the junctions, we introduce secondary vertices along edges connecting
  junctions, and we endow all secondary vertices with internal states as well. The
  internal states can describe binding, reaction, and other localized processes
  at each vertex, provided the transitions between states are linear ({\em cf.}
  \mI ). The general form of the evolution equations is 
\begin{align}
\frac{d}{dt}\bsm{N}(\bsm{X},\tau) = &
-\bsm{\lambda}\bsm{N}(\bsm{X},t)+\bsm{\lambda}\int_0^t\sum_{\bsm{X}^{\prime}\in{\mathcal N}(\bsm{X})}\bsm{W}(\bsm{X}-\bsm{X}^{\prime},t-\tau)\bsm{N}(\bsm{X}^{\prime},\tau)d\tau
\end{align}
where $ \bsm{X}$ now contains the positions and the internal state at each position,
and the $\bsm{\lambda}$ is a diagonal matrix containing the exit rate for 
 each state.  The matrix-valued transition function $\bsm{W}$ is related to
$\bsm{T}$ as described in \mI, and since we assume that  all processes have Poisson-distributed
WTDs, $\bsm{W}(\bsm{X},t) = \bsm{W}_0(\bsm{X})\delta(t)$ and $\bsm{W}_0$ simply
contains the rate constants for each possible transition between states.

The resulting stochastic process - with primary vertices, secondary vertices,
and possible transitions between internal states -- is called a multi-state
CTRW. In general such processes can be quite complicated, since they contain
many details about dynamics at a small-scale level, and the number of secondary
vertices and internal states can be large. However, we are typically interested
in assessing the evolution for longer time and space scales to determine the
overall distribution of concentration or probability of a transported substance
throughout a tissue that is many cell-lengths in size. Thus, the aim is to
simplify such processes to obtain a reaction-diffusion system with fewer
variables in which much of the local behavior has been averaged out, and in
which the macro-scale coefficients reflect the small-scale processes via
dependence of the macro-scale parameters on the microscale parameters.  The
outcome of this analysis is a reaction-diffusion system that hides much of the
micro-level details in the diffusion coefficients and the kinetic parameters,
and governs the evolution of concentrations on an entire edge.  In one space
dimension this equation can be used on any scale since the matching between
edges simply involves continuity of the concentrations and fluxes at the
junction, which by construction have no internal states. In other lattices the
same equations apply on any edge and the same matching conditions must be
satisfied, but the analysis is more complex due to the matching conditions, and
this is the subject of a future publication.

In the next section we discuss a one-dimensional example that illustrates how
micro-level parameters for several processes are reflected in macro-level
parameters for  an equation such as   Equation  \ref{1dunscalt}.

\subsubsection[Limit equations]{Derivation of the macroscale equation for a transport model}

\begin{figure}[h!]
 \centerline{
 \includegraphics[width =  3.5in]{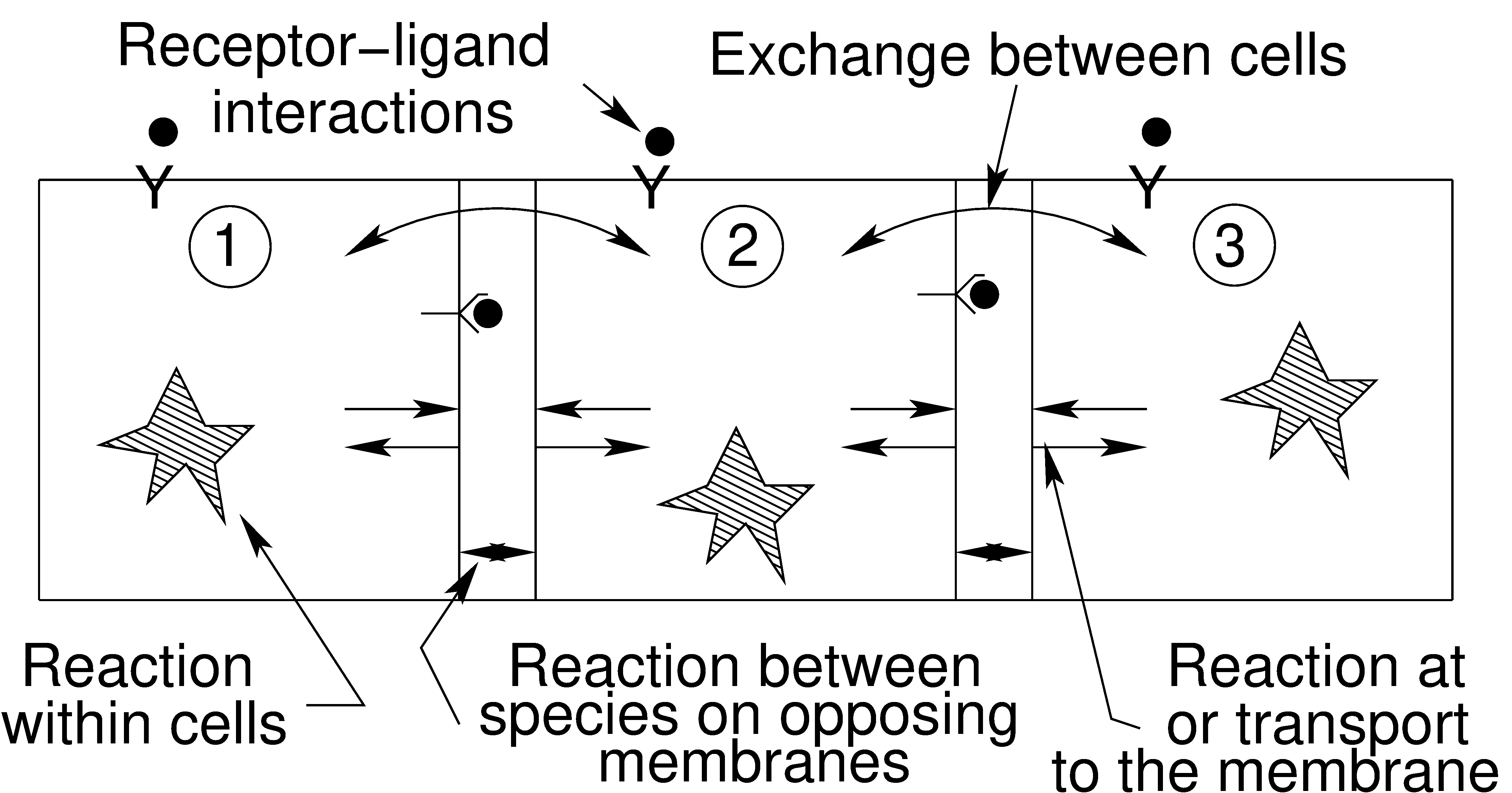}
}
 \captionsetup{margin=2cm,font=small}
\caption{ An elaboration of the transport and kinetics  processes involved in
  the wing disc.  In the analysis that follows  not all processes shown are included.
 From \protect\cite{Gou:2018:MHP} with permission.}
\label{fig:1D-disc}
\end{figure}

 Consider a line of cells,  three of which are shown in Figure
 \ref{fig:1D-disc}. separated by gap junctions\footnote{An   analysis of a  simpler system like this
   in which cells were connected by porous junctions is given in
   \cite{Othmer:1983:CMC}.} The analysis that leads to the final result is long and complex,
 and the full details are given in \mI.\, Here we simply sketch the process as an
 example to be compared with transport via cytonemes later.
We suppose that each cell is of width $L$ and separated from its neighbors by fluid junctions of width $\delta$. As shown in Fig.  \ref{fig:1D-disc}, we allow for internal reactions and
 degradation and transcytosis, and derive a macroscale equation that describes the overall
 process at large enough time and space scales. Within each cell there is
 diffusion, binding to and release from an immobile site, and degradation of the
 immobilized particle. The probability densities of states in cell $i$ evolve
 according to the following equations. 
\begin{align}
  \label{p1eqn}
\frac{\partial p_i^{(1)}}{\partial t} &= D_{m,2}\frac{\partial^2 p_i^{(1)}}{\partial x^2}\\[3pt]  
\frac{\partial p_i^{(2,1)}}{\partial t} &= D_{m,1} \frac{\partial^2 p_i^{(2,1)}}{\partial x^2}+k_- p_i^{(2,2)}-k_+p_i^{(2,1)} \\[3pt]   \label{p2eqn}
\frac{dp_i^{(2,2)}}{dt} &= -k_-p_i^{(2,2)}+k_+p_i^{(2,1)} - k_dp_i^{(2,2)}. \\[3pt]  
p_i^{(1)}(X_i,t) &= p_i^{(2,1)}(X_i,t)  \\[3pt]
p_i^{(1)}(X_i+\delta,t) &= p_{i+1}^{(2,1)}(X_i +\delta,t) \\[3pt]  
\left.D_{m,1}\frac{\partial p_i^{(1)}}{\partial x}\right|_{x=X_i} &= 
\left.D_{m,2}\frac{\partial p_i^{(2,1)}}{\partial x}\right|_{x=X_i} \\
\left.D_{m,1}\frac{\partial p_i^{(1)}}{\partial x}\right|_{x=X_i+\delta} &= 
\left.D_{m,2}\frac{\partial p_{i+1}^{(2,1)}}{\partial x}\right|_{x=X_i+\delta} &
\end{align}
Here $p_i^{(1)}(x,t)$ is the probability density that a particle is located in
the $i$th gap between the $i^{th}$ and $(i+1)^{st}$ cells, and $p^{(2,1)}$ and
$p^{(2,2)}$ represent the probabilities of mobile and bound states within a
cell.  The right-hand boundary of the $i^{th}$ cell is at $X_i$, and the gap
between cells is of length $\delta$.  The boundary conditions on $p_i^{(1)}$ and
$p_i^{(2,1)}$ represent continuity conditions on the concentrations and fluxes
for species that pass through the cell membrane, but do not bind to the membrane
itself. Finally, to connect this with the previous lattice descriptions, we
consider the right-hand cell membrane of each cell the junctions in this
system. The above equations can be viewed as the reduction of a lattice
description within cells to the reaction-diffusion equations shown.

The objective is to derive from the above equations a single
reaction-diffusion equation in which the length scale $\epsilon \equiv \delta +L$ enters as a
small quantity on the macroscopic  length scale of interest. 
A detailed analysis that  is given in \mI\, leads to 
following equation to leading order as $\epsilon \rightarrow 0$. 
\begin{align}
\begin{aligned}
\frac{\partial P(x,t)}{\partial t} &= -\frac{(1-\eta)\nu(0)}{\eta+(1-\eta)\nu^{\prime}(0)}  P(x,t) +   \frac{1}{\left(\frac{(1-\eta)}{D_{m,2}}+\frac{\eta}{D_{m,1}}\right)\left(
\eta+(1-\eta)\nu^{\prime}(0)\right)}\frac{\partial^2 P(x,t)}{\partial \xi^2}\\
&=
-\left[\frac{(1-\eta)k_dk_+(k_d+k_-)}{\left(k_d+k_- \right)^2+(1-\eta)k_-k_+} 
\right]P(x,t)+\left[\frac{1}{\left(\frac{(1-\eta)}{D_{m,2}}+\frac{\eta}{ D_{m,1}}\right)\left(\eta+(1-\eta)\left(1+\frac{k_-k_+}{(k_d+k_-)^2}  \right)\right)}\right]
\frac{\partial^2 P(x,t)}{\partial \xi^2}\\ 
&= -K_MP(x,t)+D_M\frac{\partial^2}{\partial \xi^2}P(x,t).
\end{aligned}
\end{align}
The macroscale diffusion and degradation coefficients that emerge from the
analysis are complex functions of the microscale parameters given by
\begin{equation}
K_M \equiv \frac{(1-\eta)k_dk_+(k_d+k_-)}{\left(k_d+k_- \right)^2+(1-\eta)k_-k_+} , \quad \mbox{and} \quad  D_M \equiv
\frac{1}{\left(\frac{(1-\eta)}{D_{m,2}}+\frac{\eta}{ D_{m,1}}\right)\left(\eta+(1-\eta)\left(1+\frac{k_-k_+}{(k_d+k_-)^2}  \right)\right)}
 \label{eq:Macro-Coeffs}  
\end{equation}
where $\eta \equiv \delta/L$, which remains constant in the limit $\epsilon \rightarrow 0$. Further details
  of these results are discussed in \mI, but it suffices to say that the
  derivation of the macroscale equations, though algebraically more complex,
  follows a general procedure that is applicable to many other problems. To start, we solve for the
  variables that rapidly equilibrate in terms of each other to reduce the number
  of equations. This typically yields effective WTDs that are FPT-distributions
  for transport between cells, for example, The splitting-probabilities and mean-waiting
  time are then computed, and these quantities are the only quantities that
  persist  as $\epsilon \rightarrow 0$ in the macroscale limit. There also is typically a
  discrete-Laplacian type term that reduces to a coefficient (e.g. 1/2) times
  the continuum Laplacian.

Finally, we note that in any application, $ \epsilon$ may be small but is still
nonzero. Thus, it is necessary for sufficient time to have elapsed so that the
waiting time for a single jump between adjacent cells is small compared with the
time-scales of interest. Otherwise, the discrete cellular nature of the system
will dominate and the continuum description will not be appropriate.

\subsection[Comparison of diffusion and cytonemes]{Comparison of diffusion with
  cytoneme transport in a regular hexagonal array of cells}

\subsubsection{Diffusive transport}

The previous result shows that seemingly 'simple' macroscopic diffusion may in
fact be a rather complex process when microscopic details are incorporated. In
contrast, cytoneme-based transport can be much more direct, in that a PIT
cytoneme deposits its entire load at one cell, and in this section
we derive macroscopic equations for the evolution of a population of them.  The example used focuses
on morphogen spreading in cytoneme-based transport in a hexagonal lattice of
cells that be thought of as a horizontal slice through the disc shown in Figure
\ref{discfig}(C).  Hexagonal lattices are of particular importance since many
epithelial tissues can also be approximated to first-order as a system of
hexagonally packed cells.

\begin{figure}[h!]
\begin{center}
\includegraphics[width = 0.35\textwidth]{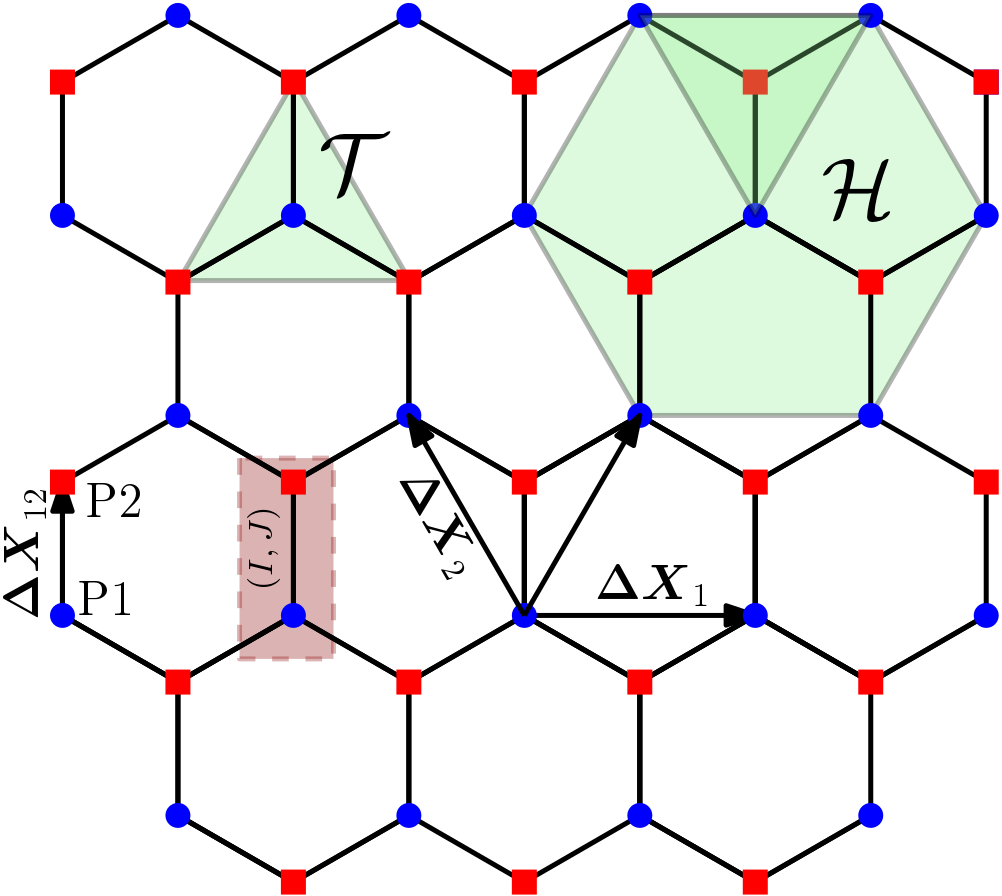}
\end{center}
\caption{ The terminology for the lattice. The 
  red points (squares) attached solely to blue points (circles) and
  vice-versa. $\Delta\bsm{X}_1$ and $\Delta\bsm{X}_2$ specify the lattice
  directions, and $\Delta\bsm{X}_{12}$ is the displacement between type I and
  type II points. The green triangle, $\mathcal{T}$ indicates the points attached to a Type I point, and the green hexagon, $\mathcal{H}$ has as vertices the Type I points that are nearest to the Type I point at its center. Modifed from \cite{Stotsky:2021:RWA}.
  \label{fig:HexLattice}}
\end{figure}

While not immediately obvious, regular hexagonal lattices can be constructed as an
alternating pattern of two topologically distinct sets of points as in Figure
\ref{fig:HexLattice}.  This connectivity structure arises since whereas all
points are connected to three neighbors, and each point has two lateral
connections, the blue (circle) points connect updwards to a red (square) point,
whereas the red points connected downards towards a blue point. Thus, the red
and blue points are topologically distinct because they exhibit a mirror-image
symmetry, but not translational symmetry. Said otherwise, while a square lattice
in the plane is self-dual and thus translation invariant, the hexagonal lattice
is not -- it's dual is a triangular lattice.
In Figure \ref{fig:HexLattice} we denote the blue points as Type I, and the red
points as Type II. It is important to note that this partitioning of the points
stems from the topology of the hexagonal lattice, and does not imply that we
must have different dynamics at the Type I and Type II points; the WTDs and all
other physical properties associated with the Type I and Type II points are
assumed equal in our models. For simplicity we only consider
junction-to-junction steps in this section -- the reader is invited to deive the
equations when there are secondary vertices and internal states.

Since the hexagonal lattice has two distinct types of points,  we have to  consider jumps from Type I
to Type II and jumps from Type II to Type I --  there are no direct jumps that take
a particle from a Type I point to another Type I point.  If we let $T_{12}$ indicate
the transitions from Type II to Type I (see the shaded triangle labeled
$\mathcal{T}$ in Figure \ref{fig:HexLattice}), and $T_{21}$ as the transitions from
Type I to Type II, we see that for diffusive transport, the two matrices are
adjoints of one another. This is a consequence of the fact that the edges of a
graph for a diffusion process are non-directed. Thus, the overall renewal
equation for this system is of the form
\[
\frac{d}{dt}
\begin{pmatrix}
P_1 \\ P_2
\end{pmatrix} 
= 
-\lambda
\begin{pmatrix}
P_1 \\ P_2
\end{pmatrix} 
+\lambda
\begin{pmatrix} \bsm{0} & T_{12} \\ T_{12}^{\ast} & \bsm{0}
\end{pmatrix}
\begin{pmatrix}
P_1 \\ P_2
\end{pmatrix} 
\]
where $T_{12}^{\ast}$ is the Hermitian adjoint of $T_{12}$. With the alternating
structure, it is convenient to solve for the Type II point probabilities in
terms of the Type I points to obtain a single equation for the Type I
points (see the shaded hexagon labeled
$\mathcal{H}$ in Figure \ref{fig:HexLattice}). For simplicity, assume that the initial condition is concentrated on
Type I points only, the general case is not difficult to obtain but involves
more book-keeping which obfuscates the underlying method. This leads to a new
equation of the form
\[
\frac{d^2 P_1}{dt^2}+2\lambda \frac{dP_1}{dt} = -\lambda^2 P_1+\frac{\lambda^2}{9}\hat{T}P_1 
\]
where the new transition matrix is of the form $\hat{T}
\equiv\bsm{T}\bsm{T}^{\ast}$. 

By enumerating all of the possible jumps of Type I points to Type I points (\cf
Figure \ref{fig:HexLattice}), one can show  that the right-hand side is
equal to $(2/3)\lambda^2 \Delta_{\mbox{tri}}P_1$ where $\Delta_{\mbox{tri}}$ is
the discrete Laplacian for a triangular lattice with spacing $\sqrt{3}L$ (see
\cite{Othmer:1971:IDP} for an explicit expression for the Laplacian).  The $\sqrt{3}L$ factor arises since
resultant length of any two adjacent edges of the hexagonal lattice is
$\sqrt{3}$ times the length of the individual edges.\footnote{The fact that the
  triangle-lattice spatial transition operator appears is result is a
  consequence of the duality between hexagonal and triangular lattices, and can
  be seen in Figure \ref{fig:HexLattice} by noting that each Type I (or Type II)
  point has six nearest-neighbor points of the same type, separated by distances
  $\pm \Delta\bsm{X}_i$ for $i=1,2$ and $\pm (\Delta\bsm{X}_1+\Delta\bsm{X}_2)$. In fact, we see that if we add points to
  the center of each hexagon, and connect adjacent hexagons, we obtain the
  triangle lattice, as shown in the darker triangle in Figure
  \ref{fig:HexLattice}). Given this observation, it is possible to consider more
  sophisticated problems with both internal and external transport by explicitly
  adding these center-points into the system and connecting them to the
  honeycomb lattice vertices via the addition of new edges.}

Now that the lattice topology has been described, it is possible to obtain a
macroscale limit equation.  In the asymptotic limit of many jumps, a particle
has generally traveled a distance many times greater than $L$, and thus, we want
to consider $L$ as a small parameter for our asymptotic analysis. We also
consider that in this case, the Laplace-transformed memory kernel $\tilde{\Gamma} =
s/\lambda^2+2/\lambda$ {\cf \cite{Stotsky:2021:RWA}.  Computing the Taylor-expansion of
the discrete Laplacian that arises here we, we obtain
\[
[\Delta_{\mbox{tri}}]P(\bsm{x}) = 3L^2 \Delta P(\bsm{x}) + O(L^4
\Delta\Delta P).
\]
Further, since we set $\lambda\sim D/L^2$, we obtain
\[
\tilde{\Gamma}=2\frac{L^2}{D} +s\frac{L^4}{D^2}.
\]
At leading order, the $L^4$ term drops out and we obtain a diffusion equation, 
\[
\frac{\partial p}{\partial t} = D\Delta p +O(L^2)
\]
While this result is still formal since we have not proven that the higher order
terms vanish, it is possible to show this result rigorously by conducting an
eigenvalue analysis to show that the eigenvalues
corresponding to fast modes (associated with differences in the concentrations
of the Type I and Type II points) vanish to $-\infty$ as $L\rightarrow 0$. The
fact that a telegrapher approximation never results is similar to the result in
\cite{Othmer:1988:MDB}. Since there is only a single time-scale (e.g. the system
is time homogeneous), the initial 'ballistic' regime, characteristic of the
telegraph's equation, disappears as $L\rightarrow 0$.  

Finally, we remark on what happens to $P_2$ in this limit. In the small-$\epsilon$
limit, a similar Taylor series expansion would show that $P_2$
reduces to a local average of $P_1$. When $P_1$ is sufficiently smooth, local averaging reduces to the
identity operator as $L$ shrinks to zero. Thus, $P_2=P_1$ to leading order at
the macroscale. 

While  we do not explore the diffusion case further, we do note that if a more
complicated model were proposed for transport along each edge, a similar
approach can be used, but as in the previous section, the diffusion coefficient
becomes for more complicated.  

\subsubsection{Transport via PIT cytonemes}

To compare diffusive transport and transport by cytonemes on a hexagonal
lattice, we derive a macroscale equation for a cytoneme transfer process of PIT
type through the lattice. We will assume that, as with the diffusive transport
described in the previous section, the cytonemes must traverse the fluid gaps
between cells, and thus move along the hexagonal lattice. However, the movement
of a tip is different in that the cytonemes extend and do not reverse direction
until they reach a target and deliver their morphogen bolus.

While each cytoneme travels without reversal, it can 'die' along an edge in that
we ignore it after it delivers its cargo to a cell.  Upon reaching a junction,
cytonemes cannot reverse, but can choose to travel along one of the two other
edges that intersect that trijunction.  While there no reversal, cytonemes could
potentially travel in a loop by making 5 right turns or 5 left turns in a
row. To eliminate this possibility, which may be biologically irrelevant, we
assume that cytonemes travel in the direction of greater $x$ in 2D whenever they
reach a junction.  Under this condition, a typical cytoneme path is as shown in
Figure \ref{fig:CytPath}.  As mentioned earlier, movement may be guided by a
signal emanating from receiver cells, which would automatically decrease the
probability of loops in populations of cytonemes.

\begin{figure}[h!]
\begin{center}
\includegraphics[width =0.3\textwidth]{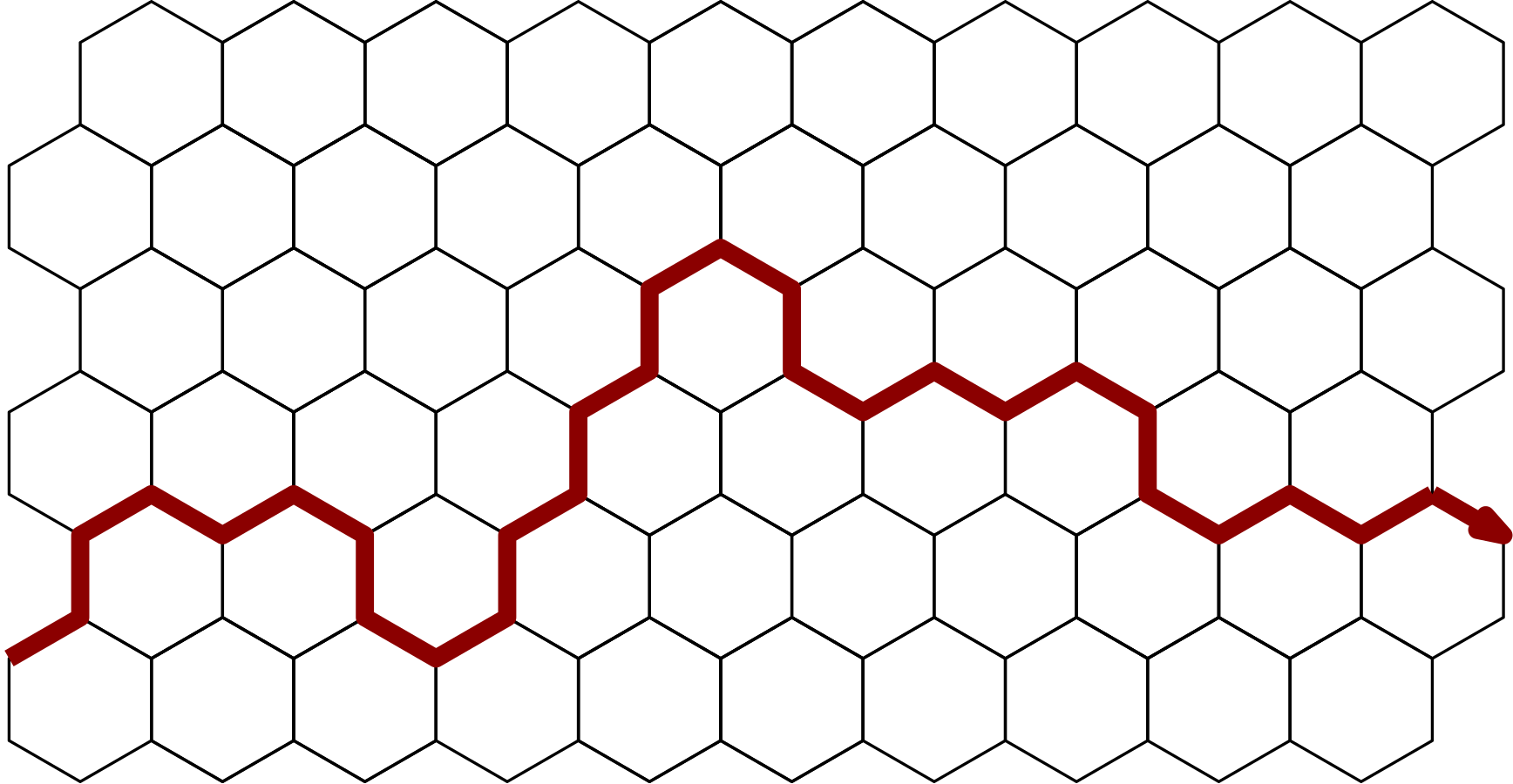}
\end{center}
\caption{The typical path of a cytoneme  under the condition that it can 
  move vertically, but always tends towards the right.
  \label{fig:CytPath}} 
\end{figure}

Since there are two topological types of
junctions, this again leads to a 2-state system. 
We will assume here that the lattice is aligned as in Figure
\ref{fig:HexLattice}, and that all cytonemes move in the $+x$-direction except
when on the vertical edges. We then have at each step a probability of $\alpha
p$ of continuing to the right, a probability of $\alpha(1-p)$ making a vertical
transition, and finally a probability of $(1-\alpha)$ for decay (see Figure \ref{fig:HexAdv}).
\begin{figure}[h!]
\begin{center}
\includegraphics[width = 0.2\textwidth]{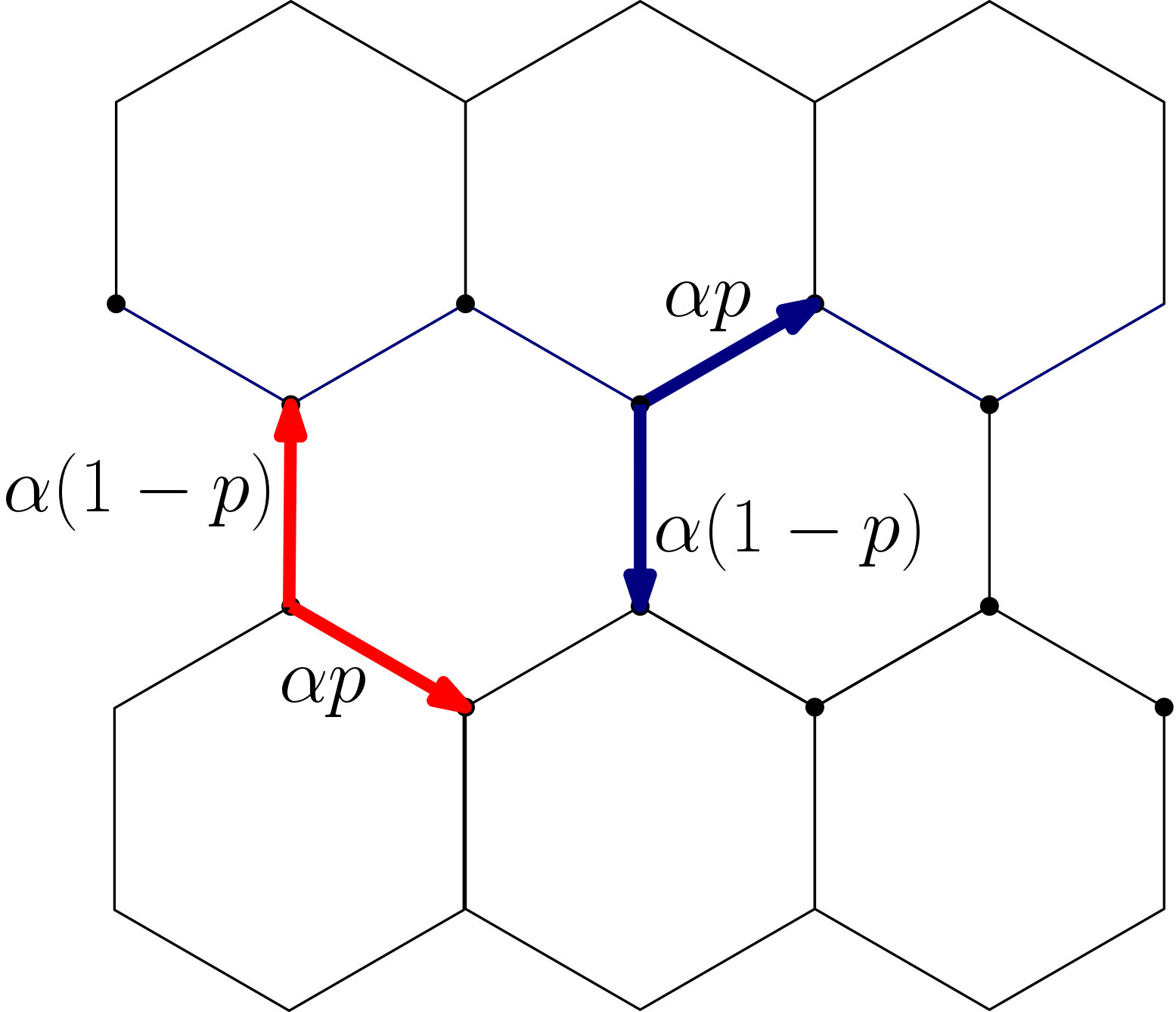}
\end{center}
\caption{Spatial jump probabilities for a cytoneme arriving departing from a
  Type I (blue) or Type II (red) point. The two parameters, $p,\alpha\in(0,1)$.}
\label{fig:HexAdv}
\end{figure} 

This system can be written at an arbitrary lattice point $\bsm{x}$ as
\begin{align}
\begin{aligned}
\frac{d}{dt}
\begin{pmatrix}
P_{I}(\bsm{x}) \\ P_{II}(\bsm{x})
\end{pmatrix}
&=
-\lambda 
\begin{pmatrix}
P_{I}(\bsm{x}) \\ P_{II}(\bsm{x})
\end{pmatrix}
+
\lambda\alpha p
\begin{pmatrix}
0 & S_{-1} \\
S_{-2} & 0
\end{pmatrix}
\begin{pmatrix}
P_{I}(\bsm{x}) \\ P_{II}(\bsm{x})
\end{pmatrix}
+
\lambda\alpha (1-p)
\begin{pmatrix}
0 & 1\\
1 & 0
\end{pmatrix}
\begin{pmatrix}
P_{I}(\bsm{x}) \\ P_{II}(\bsm{x})
\end{pmatrix}
\end{aligned}
\end{align}
where $[S_{\pm i}P](x)= P(x\pm\delta x_i)$ are shift operators that allow us to
specify connections between adjacent points on the lattice.  While it is
possible to invert this 2x2 system directly, we again reduce the system to a
single equation in order to determine the macroscopic equation that results from
this system in the limit when the edge length $L\rightarrow 0$.  We do not
expect that the distinction between Type I and Type II vertices will be
significant at the macroscale, and therefore we solve for the type II
probabilities in terms of the type I probabilities to obtain
\[
\frac{\partial^2 P_I}{\partial t^2}+2\lambda\frac{\partial P_I}{\partial t} = (\lambda^2\alpha^2(1-p)^2-\lambda^2)P_I + \lambda^2\alpha^2 p(1-p) \left[P_I(\bsm{x}-\delta\bsm{x}_1)+P_I(\bsm{x}-\delta\bsm{x}_2)\right] + \lambda^2\alpha^2p^2P(\bsm{x}-\delta\bsm{x}_1-\delta\bsm{x}_2).
\]
where $\lambda^{-1}$ is the mean time to traverse a single edge. Since the
motion is advective, $\lambda^{-1} \sim v/L$. We also set $\alpha\sim (1-k_d L)$
so that decay is not too fast to obtain a macroscale equation. The constant,
$k_d$ will be closely related to the macroscale degradation coefficient.  Then,
in the limit $L\rightarrow 0$ the equation that results is
\[
\frac{\partial P}{\partial t}  +\frac{vp}{2}\frac{\partial P}{\partial x}  = -(v k_d)P.
\]
What is particularly striking about this equation is that there is no flux term
in the $y$ direction at large scales, despite the fact that at a microscopic
scale, cytonemes can make vertical jumps. The reason is that while cytoneme
makes vertical jumps up and down, on average these vertical fluctuations cancel
out (and thus spreading along the $y$-axis is not of leading order), and thus do
not appear in the macroscopic equations.

Thus, we see that in the diffusive transport case, we obtain a
diffusion-reaction equation with coefficients that depend on the microscale
model, whereas cytoneme-transport gives a one-dimensional advection-reaction
equation. To compare these, we consider a stripe of producer cells, and for simplicity,
assume that in the diffusion case, there are reflecting boundary conditions at
the top and bottom of the tissue. If we assume uniformity in the $y$-direction,
then the 2-dimensional diffusion reduces to a one-dimensional diffusion equation
with a source at $x=0$. Under these simplifications, we can compare the
diffusion and cytoneme-based models.

To make a fair comparison, we suppose that in each case, there is a specified
flux $J$ at $x=0$ in both cases. For conciseness, we also set $v$ and $D$ equal
to the macroscale diffusion and reaction coefficients obtained above, and
$\lambda$ equal to the degradation coefficients determined above. Then, for
cytoneme-mediated transport, we have $J=vP(0,t)$, and for diffusion,
$J=-DP_x|_{x=0}$. The resulting solutions to this 1-dimensional problem are then
given as
\begin{align}
\begin{aligned}
P(x,t) &= \frac{J}{v}H(t-x/v)e^{-\frac{\lambda}{v} x} \\ 
P(x,t) &=
\frac{J}{2\sqrt{\lambda D}}\left[e^{-\sqrt{\lambda/D}x}
  \mbox{Erfc}\left(\frac{x-2t\sqrt{D\lambda}}{2\sqrt{Dt}} \right)-
  e^{\sqrt{\lambda/D}x}\mbox{Erfc}\left(\frac{x+2t\sqrt{D\lambda}}{2\sqrt{Dt}}
  \right) \right].
\end{aligned}
\end{align}
Several observations can be made from these solutions. Firstly, when
$D=1\mu m^2/\mbox{sec}$ and $v=1\mu m /\mbox{sec}$ with $\lambda = 1/\mbox{sec}$,
both solutions are equivalent in the limit $t\rightarrow \infty$ suggesting that
in biological situations where the diffusivity, advective velocity, and reaction
rates are of roughly equal importance, either strategy could be applied in
theory. On the other hand, the dynamics are quite different. Under diffusion,
the concentration distribution gradually increases everywhere in space, whereas
under advection, the concentration distribution is unfolded over time, but once
it is set, does not change.

When $\lambda$ is allowed to vary, we see that for $\lambda<1$, diffusion leads
to a broader distribution relative to advection, whereas when $\lambda>1$, the
opposite occurs. Both situations have potential roles biologically, for
instance, if it is important that distant cells be reached, then the strategy
that yields a greater extent of travel might be more useful, whereas, if cells
are detecting the gradient of a substance, than the shorter, and hence more
sharply peaked distribution might be more effective at achieving biological
goals.

\section{Discussion}

Herein we have developed deterministic and stochastic models of transport
by cytonemes, both when the receiver cell searches for a producer cell, and when
the producer cell extends cytonemes to deliver cargo to receivers.
 In simple models analytical solutions are attainable, and the
complete spatiotemporal dependence of the morphogen concentration has been
obtained. A key quantity that arises is the rate $\mu/v$ of the cytoneme halting
rate compared with its velocity. This parameter controls the steepness of the
cytoneme concentration field, and hence how far morphogen can spread.

We then discussed the generalization of the simple models to include stochastic
cytoneme generation rates, and discussed some analytical results concerning the
statistics of morphogen received at a given cell in this case. A key point in
these models is that since cytonemes are presumed not to interact amongst
themselves, the position of the receiving cell shows up merely as a parameter in
the models. It would be of interest to ascertain experimentally whether or not
nearby cytonemes can interact, and whether this has implications for morphogen
transport.

We next considered a stochastic simulation approach for more comlex models.
While many of the features of the simpler model is present in this case, the
basic stochastic models are easy to extend to include features such as feedback
mechanisms whereby cells turn off cytoneme generation after receiving enough
morphogen. There are many further extensions that can be applied here - notably,
it is straightforward to include more complicated morphogen extension and
retraction steps, and the possibility of a variable morphogen bolus size and  even
the ability to consider dynamic morphogen transport along cytonemes that are
attached to a producer cell.

Finally, we discussed how cell-level models that include details of the
diffusive or cytoneme-mediated transport processes can be upscaled to obtain
tissue-level macroscale equations. These macroscale equations take on much
simpler forms, but have coefficients that depend in complex ways on the
microscale details of the transport procedures. At this level, we discussed a
simple comparison between diffusive transport and cytoneme-mediated
transport. We found that either model can yield the same
distribution of morphogen after some time, but that the dynamics of the approach
to the longer-term morphogen distribution differ between the modes. This
suggests that understanding the importance of diffusion and cytoneme-based
transport to morphogen spreading may require not only analysis of 'steady-state'
distributions of morphogen, but also the dynamics of the morphogen concentration
field over time.  

\section{Acknowledgments}
This work was supported in part by NSF Grants DMS \# 178743 and 185357.

\bibliographystyle{./AIMS/AIMS}
\bibliography{a-l-3-3-22,m-z-3-3-22,./bibfiles/wingdisc3,./bibfiles/AddtlRefs,./bibfiles1/RandomWalks}

\section{Appendices}
\subsection{Computation of FPT for cytonemes with a resting phase}

Let us consider the computation of the FPT for the model of cytonemes extending
and retraction with a resting phase. This is an example of an alternating
renewal process, and the solution for the FPT is computable via the use of
Laplace-transforms.

The governing equations for this model are described in Section \ref{subsec:PIT}
and Equations \eqref{eq:RestingPhaseModel}. After Laplace transformation, and
assuming an initial condition that a cytoneme is generated at $x_0$ at $t=0$, we
obtain an algebraic system for the Laplace-transformed quantities
\begin{align}
\begin{aligned}
s\tilde{p}_e-v\tilde{p}_{e,x} &= -\lambda_m \tilde{p}_e + \lambda_r \tilde{r}_1
+\frac{1}{v}\delta(x-x_0) \\ s\tilde{r}_1 &= \lambda_m \tilde{p}_e-\lambda_r
\tilde{r}_1\\ s\tilde{P}_a &= v\tilde{p}_e(0,s)-\mu_d\tilde{P}_a
\\ s\tilde{p}_r+v\tilde{p}_{r,x} &= -\lambda_m \tilde{p}_r + \lambda_r
\tilde{r}_2 + \delta(x)\frac{\mu_d}{v}\tilde{P}_a\\ s\tilde{r}_2 &= \lambda_m
\tilde{p}_r -\lambda_r \tilde{r}_2 \\
\end{aligned}
\end{align}
For the FPT problems, we are interested in the quantity $vp_r(x_0,t)$, which can
be found from above as
\[
\tilde{f}(s|x_0) =
\frac{\mu_d}{s+\mu_d}e^{-\frac{2x_0}{v}\left(s+\lambda_m-\frac{\lambda_m\lambda_r}{s+\lambda_r}\right)}
\]
To invert this, first note that the product of Laplace-transform domain
quantities leads to their convolution in real-time and that the inverse
transform of $e^{-a s}\tilde{g}(s)$ leads to step functions, $H(t-a)g(t-a)$
where $g(t)$ is an arbitrary function whose Laplace-transform is sufficiently
well-behaved. With these result, let us consider just the inverse transform of
\[
e^{-\alpha/(s+\lambda_r)}.
\]
We can further use the Laplace-transform shift formula and just consider
$e^{-\alpha/s}$. While the inversion integral is involved, it can be shown (or
looked up in a table), that the inverse transform is
\[
\sqrt{\frac{\alpha}{t}}I_1(2\sqrt{\alpha t})+\delta(t)
\]
where $I_1(\alpha)$ is the modified Bessel-function of the first kind of order
1.  Applied to our case, we obtain an inverse
\[
f(t|x_0) = \int_0^t \left[\mu_d
  e^{-\mu_d(t-\tau)}H\left(\tau-\frac{2x_0}{v}\right)\left[e^{-\left(\tau+\frac{x_0}{v}\right)\lambda_m}
    \sqrt{\frac{c}{\left(\tau-\frac{2x_0}{v}\right)}} I_1\left(2\sqrt{c
      \left(\tau-\frac{2x_0}{v}\right)
    }\right)+\delta(\tau-\frac{2x_0}{v})\right]\right] d\tau
\]
with $c=2\lambda_m\lambda_r x_0/v$.  The convolution integral that results does
not appear to have any obvious simplification, but could be numerically
evaluated if needed. On the other hand, from the Laplace-transformed result, it
is straightforward to compute moments with respect to $t$ by differentiating
$\tilde{f}(s)$ at $s=0$.  For instance,
\[
\langle t\rangle = \frac{1}{\mu_d} +
\frac{2x_0}{v}\left(1+\frac{\lambda_m}{\lambda_r}\right)
\]
Notice that the two terms, $1/\mu_d$ and the velocity dependent term essentially
account for 1) the cytoneme to be attached to the producer cell, and 2) the time
it takes to go out and come back.

The variance can be computed by taking the second derivative at $s=0$ and
subtracting $\langle t\rangle^2$. This leads to
\[
\sigma^2(t) = \langle \left(t-\langle t\rangle \right)^2\rangle =
\frac{1}{\mu_d^2}+\frac{4x_0}{v}\frac{\lambda_m}{\lambda_r^2}
\]
which again, has components due to the variance of waiting while attached, and
of the time it takes to travel to and from the producer cells.

\subsection{Simplification of double sum for the composite process}
We found in Section \ref{subsec:ComposRand} that the distribution function for
sums of Bernoulli variables that are themselves dependent on a Poisson process
results in a rather simple form, namely that the the resulting distribution is
exponentially distributed with a time-dependent rate-parameter.

Here, we show how this may be computed starting from Equation
\eqref{eq:BernoulliDist}, reproduced below
\[
\mathbb{P}\left[\sum_{i=1}^{N(t)}R(t|t_i)-N_r\geq 0 \right] =
\sum_{n=0}^{\infty} p_n(t)\mathbb{P}_n\left[\sum_{i=1}^nR(t|t_i)>N_r\right].
\]
We assume that each $R_i(t|t_i)$ is itself a random variable, and that all the
$R$'s are i.i.d. and take on values of $0$ and $1$ with some probability for
each $t>t_i$.  At time $t$, let $r(t|0)dt$ be the probability that $R$ goes from
$0$ to $1$ in the interval $(t,t+dt)$.  As an example, this could be the
probability that a cytoneme which started at $t=0$ returns to the cell it
started at around time $t$. Since the $t_i$ are also random times, we can define
a density for the probability of obtaining $N_r$ successes (e.g. $R$ goes from 0
to 1) in the intervals $(t_1,t_1+dt)$, $(t_2,t_2+dt)$, ..., $(t_n,t_n+dt)$. This
is equal to the sum over all possible combinations of $j$ out of $n$ $R_i$'s
becoming successful at $t$. Since the $R_i$'s are all i.i.d., this can be
written as
\[
P(t,N_r|n,\{t_1,\dots,t_n\}) = \sum_{A\in V_{N_r,n}}\prod_{i\in
  A}r(t|t_i)\prod_{j\in A^c}(1-r(t|t_j))
\]
where $V_{N_r,n}$ is the set of all combinations of $N_r$ elements of
$\{t_1,\dots,t_n\}$, $A = \{t_{a_1},\dots,t_{a_{N_r}}\}$ is an element of
$V_{N_r,n}$, and $A^c$ is the complement of $A$. This appears quite complicated
at first glance since each of the Bernoulli variables has a potentially distinct
success probability at $t$ since each $t_i$ is generally distinct from the
others. However, we next want to take the expectation over all of the $t_i$ to
find the probability of having $N_r$ successes at any combination of times prior
to $t$. To integrate, we note that due to independence, in each term of the sum,
each $t_i\in\{t_1,\dots,t_n\}$ appears precisely once, and upon taking the
expectation of any particular element, $A\in V_{N_r,n}$, we obtain
\begin{align}
\begin{aligned}
\mathbb{E}_1\left[\prod_{i\in A}r(t|t_i)\prod_{j\in A^c}(1-r(t|t_j))\right]
&=\frac{1}{t^n} \underset{n\mbox{ times}}{\underbrace{\int_0^t\dots
    \int_0^t}}\left[\prod_{i\in A}r(t|t_i)\prod_{j\in A^c}(1-r(t|t_j))\right]
dt_1dt_2\dots dt_n \\ & = \prod_{i\in A}\mathbb{E}_1[r(t|\cdot)]\prod_{j\in
  A^c}(1-\mathbb{E}_1[r(t|\cdot)] =
\mathbb{E}_1[r(t|\cdot)]^{N_r}(1-\mathbb{E}_1[r(t|\cdot)]^{n-N_r}
\end{aligned}
\end{align}
For short-hand, define $\bar{r}(t) = \mathbb{E}_1[r(t|\cdot)]$. The cardinality
of $V_{N_r,n}$ is the set of all possible combinations containing $N_r$
successes out of $n$ tried, which is just $\binom{n}{N_r}$, thus leading to
\[
\mathbb{E}\left[P(t,N_r|n\{t_1,\dots,t_n\})\right] = P(t,N_r|n) = \binom{n}{N_r}
\bar{r}(t)^{N_r}(1-\bar{r}(t))^{n-N_r}.
\]
In general, we are interested in the probability of $\sum_{i=1}^n R_i(t|t_i)\geq
N_r$, thus we must sum over $j=N_r$ to $j=n$, e.g.
\[
\sum_{j=N_r}^n \binom{n}{j} \bar{r}(t)^{j}(1-\bar{r}(t))^{n-j}.
\]
and finally compute the expectation over $n$. For a Poisson process with
constant rate, this leads to
\[
\mathbb{P}\left[\sum_{i=1}^{N(t)}R(t|t_i)-N_r\geq 0
  \right]=\sum_{n=0}^{\infty}e^{-\lambda t}\frac{(\lambda t)^n}{n!}
\sum_{j=N_r}^n\binom{n}{j} \bar{r}(t)^{j}(1-\bar{r}(t))^{n-j}
\]
To compute this sum, it is best to interchange the order of summation between
$j$ and $n$. Note that in each term it is always true that, $n\geq j$. After
interchanging the order and finding the appropriate bounds on the sums, we
obtain

\[
\mathbb{P}\left[\sum_{i=1}^{N(t)}R(t|t_i)-N_r\geq 0
  \right]=\sum_{j=N_r}^{\infty}\sum_{n=j}^{\infty}e^{-\lambda t}\frac{(\lambda
  t)^n}{n!} \binom{n}{j} \bar{r}(t)^{j}(1-\bar{r}(t))^{n-j}
\]
Further simplification is done by defining $m=n-j$ yielding
\[
\mathbb{P}\left[\sum_{i=1}^{N(t)}R(t|t_i)-N_r\geq 0 \right]=e^{-\lambda t
  \bar{r}(t)}\sum_{j=N_r}^{\infty}\frac{(\lambda t \bar{r}(t))^j}{j!} = 1 -
e^{-\lambda t \bar{r}(t)}\sum_{j=0}^{N_r-1}\frac{(\lambda t \bar{r}(t))^j}{j!}
\]
We see that this happens to be the cumulative sum for a exponentially
distributed discrete variable. Furthermore, it is known that the sum can be
written in 'closed' form in terms of the upper incomplete Gamma-function:
$\Gamma(N_r,\lambda t\bar{r}(t))$,

\[
\mathbb{P}\left[\sum_{i=1}^{N(t)}R(t|t_i)-N_r\geq 0 \right] =
\frac{1}{(N_r-1)!}\int_{\lambda t\bar{r}(t)}^{\infty}z^{N_r-1}e^{-z}dz.
\]
Note that regardless of how complex the dynamics of $R_i(t|t_i)$ are, this
results and is a consequence of the independence of the motion of all of the
cytonemes. It is of future interest to determine whether similar results may be
obtainable when there is some dependence between cytonemes either due to spatial
interactions, or when the cytoneme generation process is not a Poisson process.

\end{document}